\documentclass[11pt]{article}

\usepackage{graphicx,epsfig}
\usepackage{multicol}
\usepackage{amsmath,amssymb,cite,color,hyperref}
\newcommand{\be}{\begin{equation}}
\newcommand{\ee}{\end{equation}}
\newcommand{\bea}{\begin{eqnarray}}
\newcommand{\eea}{\end{eqnarray}}
\newcommand{\bi}{\begin{itemize}}
\newcommand{\ei}{\end{itemize}}
\newcommand{\ben}{\begin{enumerate}}
\newcommand{\een}{\end{enumerate}}

\newcommand{\lc}{\left[}
\newcommand{\rc}{\right]}
\newcommand{\lp}{\left(}
\newcommand{\rp}{\right)}

\def\frac#1#2{{{#1}\over {#2}}}
\def\gsim{\mathrel{\rlap{\lower4pt\hbox{\hskip1pt$\sim$}}
    \raise1pt\hbox{$>$}}}         %greater than or approx. symbol
\def\lsim{\mathrel{\rlap{\lower4pt\hbox{\hskip1pt$\sim$}}
    \raise1pt\hbox{$<$}}}         %less than or approx. symbol

\newcommand{\draft}[1]{}

\definecolor{grey}{rgb}{0.5,0.5,0.5}

\def\ttbar{t\bar{t}}
\begin{document}

\title{\textbf{\sc Cross Section Ratios between different CM energies at the
  LHC: opportunities for precision measurements and BSM sensitivity}}

\author{
Michelangelo L. Mangano$^a$ and Juan Rojo$^{a}$\\[5pt]
~\\
\small
$^a$ PH Department, TH Unit, CERN, CH-1211 Geneva 23, Switzerland\\[5pt]
\small
E-mail: michelangelo.mangano@cern.ch, juan.rojo@cern.ch}

%\begin{document}

\maketitle
\date{}

\vspace{1cm}
\begin{abstract}
The staged increase of the LHC beam energy provides a new class of
interesting observables, namely ratios and double ratios of cross
sections of various hard processes. The large degree of correlation of
theoretical systematics in the cross section calculations at different
energies leads to highly precise predictions for such ratios. We
present in this letter few examples of such ratios, and discuss their
possible implications, both in terms of opportunities for
precision measurements and in terms of sensitivity to Beyond
the Standard Model dynamics.
\end{abstract}
\vfill
%\vspace{5cm}
\begin{flushleft}
CERN-PH-TH/2012-159\\
\today
\end{flushleft}

\section{Introduction}

\label{sec:intro}

The excellent performance of the LHC accelerator complex and of the experiments has allowed, in just a matter of a couple of years, for very precise cross section measurements covering a broad range of hard final states, such as 
jets~\cite{CMS:2011mea,Chatrchyan:2011qta,Collaboration:2011fc}, 
top quarks~\cite{atlastop1,atlastop2,atlasstop,cmstop1,cmsstop}, 
electroweak gauge 
bosons~\cite{Aad:2011dm,Aad:2011yn,cmsweasy,
Chatrchyan:2011jz,Chatrchyan:2011wt,Aaij:2012vn}, 
direct photons~\cite{Aad:2011tw,Chatrchyan:2011ue}, 
and associated production thereof~\cite{CMSWc,ATLAS:2012ar,atlaswjets}.
 When these results are compared to the available
precise theoretical 
calculations~\cite{Maestre:2012vp}, 
improved information can be inferred
on the structure of the proton~\cite{Watt:2011kp,d'Enterria:2012yj} 
and of the Standard Model (SM)
parameters, and indications of possible departures from the SM itself
can be detected. Given the rich statistics of the LHC data, this
programme is typically limited in its potential only by the nature and
size of the systematic 
uncertainties that accompany both the measurements and the
theoretical calculations. To the extent that such uncertainties can be
correlated among different processes, it is possible to define
combinations of various observables that can be calculated, and
possibly measured, with a higher degree of precision.

Several works in the past have already introduced key ideas to define
a programme of precision cross section measurements at the LHC. For
example, refs.~\cite{Dittmar:1997md,Khoze:2000db,Krasny:2007cy} introduced and 
explored the use of precisely known
Drell-Yan and QED processes to extract indirectly the absolute LHC
luminosity, and to correlate the PDF systematics in 
cross sections of different processes. More recently,
refs.~\cite{Krasny:2011hr,Krasny:2011dj} discussed the precision and
discovery potential of
comparing rate measurements performed at different beam energies, and
with different beam types. 

The goal of this letter is to provide an up-to-date quantitative
estimate to the intrinsic theoretical precision of ratios and double
ratios of cross sections for different processes, measured at different
beam energies, using the most accurate perturbative knowledge and the
latest PDF sets. We focus on observables that are already routinely
measured by the LHC detectors, and for which the performance has been
established. Depending on which ratios one is willing to take from
theory as benchmarks, this precision can be used to correlate
luminosity measurements at different energies, to validate
experimental measurements of specific cross sections, to test and
improve PDF fits, and even to probe the existence of underlying
BSM phenomena.

A typical example of the quantities that have been discussed in the literature 
is the ratio of the cross section for a given
process $X$ by the production rate of
$Z$ bosons. The latter is the most precisely determined rate in hadronic
collisions, both theoretically and experimentally. Rescaling the rate
of $X$ to the
$Z$ cross section removes entirely the experimental uncertainty on the
LHC luminosity, and could also lead to a further reduction of the
theoretical sensitivity to the parton density functions (PDFs) that
parametrize the proton quark and gluon structure. This is true, for
example, of the ratios $\sigma(Z+{\rm jet})/\sigma(Z)$, or
$\sigma(W)/\sigma(Z)$. However, the theoretical
systematics, such as the scale uncertainty, are generally totally
uncorrelated between the process $X$ and $Z$ production, and cannot be
reduced by taking this ratio.

In this letter we explore the nature of the correlations among
theoretical
systematics, and the precision potential, of measurements
taken at different LHC beam energies. This is motivated by the staged
approach of the LHC to higher energies, with large sets accumulated,
so far, at $\sqrt{s}=7$~TeV, and expected this year at
$\sqrt{s}=8$~TeV, and following 2014 at $\sqrt{s}\sim 14$~TeV.

The key observation is that, for a given process, the calculation of 
cross sections at different energies requires correlated values of the
various input parameters, such as masses, $\alpha_S$,
PDFs and scales. When these parameters are varied simultaneously at the two
energies, the cross section ratio is much less sensitive to their
variation. When comparing this prediction to data, three things can
happen:
\begin{itemize}
\item if the residual theoretical systematics is dominated by PDFs (as
  will be the case in several of the examples shown below), and is larger than the
  achievable experimental precision, the data versus theory comparison can
  be used to improve the PDF determination;
\item the theoretical systematics could be small enough that,
  with a comparable experimental precision, the measurement becomes
  sensitive to possible contributions from physics beyond the SM (BSM);
\item when residual systematic errors are small enough, and no 
BSM contribution is within reach, one can use the cross section ratios
as standard candles for luminosity measurement and cross calibration; for
example, to correlate the luminosity measurement between two beam
energies and between two different experiments.
\end{itemize}
In all cases, the improved theoretical precision provides the
experiments with an important diagnostic tool for the validation of the
analyses at different energies, and a benchmark to be used for a
possible reduction of some of the experimental uncertainties.

We give in this paper a few examples to illustrate the above
points. Several other cases of interest can be considered, but a
conclusive assessment of whether these proposals are indeed useful
will require an in-depth analysis of the challenge posed by
correlating the experimental systematics at different energies.
Experimental cross section systematics typically fall in three
categories: efficiencies/acceptances, energy scales and absolute
luminosity. The first arises from a mixture of purely experimental
effects and theoretical modeling: in the case of leptonic final
states, such as $W$ or $Z$ bosons measurements, these uncertainties
can be brought to a (sub)percent level, and further reduced in ratios.
In the case of jets, the energy scales dominate the uncertainties, at
the level of up to 10-20~\%. While the event structure varies at
different CM energies and running conditions (e.g. due to pile-up of
multiple $pp$ interactions), it is likely that a large fraction of
these systematics can be correlated. 

The direct measurement of the LHC absolute luminosity has improved
significantly over the last year, and is now estimated to attain the
level of 2\%, a systematics which is however uncorrelated among
different CM energies~\cite{lumidays}. The simultaneous measurement of
total $pp$ cross section and luminosity, at the level of few percent,
can be obtained, by using the optical theorem, by the
TOTEM~\cite{totem,Antchev:2011vs} and ALFA~\cite{alfa}
detectors. New detector concepts have been
proposed~\cite{Krasny:2006xg,Krasny:2010vj,Krasny:2011hr}, capable of
measuring and fully exploiting the precisely known cross section for
exclusive electromagnetic production of dilepton pairs ($pp\to pp
\ell^+\ell^-$)~\cite{Budnev:1973tz,Shamov:2002yi,Khoze:2000db}. These
could improve the absolute luminosity measurement to 1\%, and the
relative luminosity at different energies, and for different hadronic
beam types, to 0.1\%~\cite{Krasny:2011hr}.  Nevertheless, the only way
to further reduce the luminosity uncertainty in the ongoing runs, in
fact to fully remove it, is to take cross-section ratios for different
processes.

The outline of this letter is as follows. On Sect.~\ref{sec:settings}
 we provide more details on the theoretical framework
for our calculations and present  explicit
quantitative results relative to production of electroweak gauge bosons, top
quarks, Higgs boson and jets.
Then in Sect.~\ref{sec:lumi} 
we discuss how to understand the generic dependence
of cross section ratios in terms of ratios of PDF
luminosities. In Sect.~\ref{sec:bsm}
we discuss how the possible sensitivity to BSM
contributions might be enhanced in cross section ratios,
and then we conclude.

\section{Theoretical systematic errors in cross section ratios}
\label{sec:settings}

The main focus of the present paper will be the 
 ratio of cross sections for a final state $X$ between
different LHC center of mass energies $E_{1,2}=\sqrt{s_{1,2}}$: 
\be R_{E_2/E_1}(X)\equiv
\frac{\sigma(X,E_2)}{\sigma(X,E_1)} \; ,
\ee 
with $E_i=7,8$ or 14 TeV.\footnote{There is also data at $\sqrt{s}=2.76$  TeV,
used mostly for $pp$ benchmarks in $PbPb$ measurements, but experimental
uncertainties are larger and thus we will not consider this case here.}   
In view of the cancellation of the luminosity, we shall also consider
double ratios of cross sections:
\be
R_{E_2/E_1}(X,Y)\equiv
\frac{\sigma(X,E_2)/\sigma(Y,E_2)}{\sigma(X,E_1)/\sigma(Y,E_1)}
\equiv
\frac{R_{E_2/E_1}(X)}{R_{E_2/E_1}(Y)} \; . 
\ee

In this work we consider two classes of observables. 
First of all we consider inclusive cross sections 
for electroweak gauge boson production, top quark pair
production, and Higgs boson production in the gluon fusion channel.
Next, we consider more differential distributions, in 
particular the top quark pair cross section above a certain
threshold in the invariant mass of the $t\bar{t}$ pair
and inclusive jet production in a given range of $p_T$
and rapidity.
For each of these processes we have evaluated all the
relevant  associated systematic theoretical uncertainties due to:
\begin{itemize}
\item Parton Distribution Functions
\item Higher perturbative orders
\item Values of $m_t$ and $\alpha_s\lp M_Z\rp$
\end{itemize}
To be more precise, we have used the following codes
and settings for cross section computations:
\begin{itemize}
\item Electroweak gauge boson production has been computed at NNLO
  using the {\tt Vrap} code~\cite{Anastasiou:2003ds}. The central
  scale is $Q^2=M_V^2$.
\item Top quark pair production has been computed at NLO+NNLL with the
  {\tt top++} code~\cite{Czakon:2011xx}\footnote{We did not include
    the latest development of the calculation of the complete NNLO
    corrections to the $q\bar{q}\to t\bar{t}$ production, documented
  in~\cite{Baernreuther:2012ws}. Their effect would be to further
  reduce slightly the theoretical scale systematics.}.  The central
    scale is $Q^2=m_t^2$.  The settings of the theoretical
    calculations are the default ones in Ref.~\cite{Cacciari:2011hy}.
\item Higgs boson production cross sections in the gluon fusion
  channel have been computed at NNLO with the {\tt iHixs} code~\cite{Anastasiou:2011pi}.  The central scale has been taken to be
  $Q=M_H/2$, to simulate the effects of 
NNLL resummation~\cite{deFlorian:2009hc}.
\item Top quark pair production with a lower cut on the top-antitop
  quark pair invariant mass of 1 TeV and 2 TeV has been computed at
  NLO with the {\tt MCFM6.2} code~\cite{Campbell:2002tg,Campbell:2004ch}, and cross-checked with the
       {\tt MNR} code~\cite{Mangano:1991jk}.  The central scale has
       been taken to be $Q^2=M^2_{t\bar{t}}$, the invariant mass of
       the top-antitop pair. This is a suitable choice of scale since
       it is of the same order of magnitude of the typical hard scales
       involved in these processes. We verified that the scale
       systematics for cross section ratios is consistent with the
       alternative choice of $Q^2=m_t^2+\langle p_T^2\rangle$, where
       $\langle p_T^2\rangle = (p_{T,t}^2+p_{T,\bar{t}}^2)/2$.
\item Inclusive jet production with a lower cut in the
  transverse momentum of the jet of 1 TeV and 2 TeV in the region
$|\eta|\le 2.5$  has been computed at NLO 
 with a modified version of the {\tt EKS} jet production
  program~\cite{PhysRevD.46.192}. The calculation uses the anti-$k_T$
  algorithm with $R=0.6$, with the scale in each event set
equal to the $p_T$ of the
  hardest jet in the event. As cross-checks, we have also
computed inclusive jet cross sections with the {\tt FastNLO} 
program~\cite{Kluge:2006xs,Wobisch:2011ij}, 
based on {\tt NLOjet++}~\cite{Nagy:2003tz}, for 7 and 8 TeV, for a fine
binning in the transverse momentum of the jet and in the central
region with $|\eta|\le 0.5$.
\end{itemize}

The choice of PDF sets, and the values of the SM parameters and
the calculation of theoretical systematics adopted
in our computation are the following:
\begin{itemize}
\item The reference PDF set is the NNPDF2.1 NNLO set~\cite{Ball:2011mu,Ball:2011uy}. For cross-checks, and to gauge
the sensitivity with respect the choice of PDF
set, we will use in addition the MSTW2008nnlo~\cite{Martin:2009iq} and ABKM09 NNLO~\cite{Alekhin:2009ni} PDF sets.
\item For all  processes, renormalization and factorization
  scales have been varied in the range $0.5 \le \mu_R/Q,\mu_F/Q\le 2$,
  with the constraint that $0.5 \le \mu_R/\mu_F\le 2$, to avoid
  artificial large logarithms of scale ratios. For ratios of different
  observables, scale variations in the numerator and in the denominator
  are taken as uncorrelated, and thus added in quadrature, 
 except for $W$ and $Z$ ratios where scale
  variations are taken as fully correlated between numerator and
  denominator.  The scale uncertainty $\delta_{\rm scale}$ is defined
  as the maximum (minimum) difference between the result at the
  central scale and the results varying the scales in the above range.
\item PDF uncertainties are computed directly on the cross section
  ratios, keeping track of all the PDF induced correlations between
  numerator and denominator.
\item The reference value for the strong coupling has been taken to be
  $\alpha_s\lp M_Z^2\rp=0.119$, and a conservative uncertainty of
  $\delta_{\rm \alpha_s}=0.002$ at the 68\% CL is assumed~\cite{Bethke:2009jm}.  The correlations between PDFs and $\alpha_s$
  are consistently taken into account, using the NNPDF sets
with varying  $\alpha_s$ as described in~\cite{Demartin:2010er,Ball:2010de}
\item The reference value of the top quark mass is taken to be
  $m_t=173.3$ GeV~\cite{0954-3899-37-7A-075021},
 with a conservative uncertainty of $\delta_{m_t}=2$
  GeV. We have verified that the dependence of the cross section ratios
on the top quark mass is very small, for example for the 8 over 7 TeV
ratio it is at most 1 permille, much smaller than any other theory
systematics, and this is the same for all other cases studied.
Therefore in the following we will not provide the explicit
contribution of $\delta_{m_t}$ to the total theory systematics,
since is always negligible as compared to PDF, scale and
 $\alpha_s$ uncertainties.
\item We assume a Standard Model Higgs with mass $m_H=125$ GeV. We
  verified that the impact on our results of a possible
  $\delta_{m_H}=2$ GeV uncertainty on
  its mass is negligible.
\end{itemize}

%%%%%%%%%%%
\begin{table}[t]
\centering
\small
 \begin{tabular}{c|c|c|c|c}
 \hline
 Cross Section & $R^{\rm th,nnpdf}$ & $\delta_{\rm PDF}$(\%)  & $\delta_{\alpha_s}$ (\%)&$\delta_{\rm scales}$ (\%)\\
 \hline
 \hline
$t\bar{t}/Z$                                       &       1.23 &$\pm$   0.4 & $  -0.2$ --    0.2 & $  -0.2$ --    0.3\\  
$t\bar{t}$                                         &       1.43 &$\pm$   0.3 & $  -0.2$ --    0.2 & $  -0.1$ --    0.3\\  
$Z$                                                &       1.16 &$\pm$   0.1 & $  -0.0$ --    0.1 & $  -0.1$ --    0.1\\  
$W^+$                                              &       1.15 &$\pm$   0.1 & $  -0.0$ --    0.1 & $  -0.1$ --    0.1\\  
$W^-$                                              &       1.17 &$\pm$   0.1 & $  -0.0$ --    0.1 & $  -0.1$ --    0.1\\  
$W^+/W^-$                                          &       0.98 &$\pm$   0.1 & $   -0.0$ --    0.0 & $  -0.0$ --    0.0\\  
$W/Z$                                              &       0.99 &$\pm$   0.0 & $  -0.0$ --    0.0 & $  -0.0$ --    0.0\\  
$ggH$                                              &       1.27 &$\pm$   0.2 & $  -0.0$ --    0.1 & $  -0.2$ --    0.2\\  
$t\bar{t}(M_{tt}\ge 1~{\rm TeV})$                  &       1.81 &$\pm$   0.8 & $   -0.0$ --    0.3 & $  -0.6$ --    0.5\\  
$t\bar{t}(M_{\rm tt}\ge 2~{\rm TeV})$              &       2.80 &$\pm$   3.2 & $  -0.6$ --    0.3 & $  -0.0$ --    1.4\\  
$\sigma_{\rm jet}(p_{T}\ge 1~{\rm TeV})$           &       2.30 &$\pm$   1.0 & $   -0.0$ --    0.5 & $  -0.4$ --    1.0\\  
$\sigma_{\rm jet}(p_{T}\ge 2~{\rm TeV})$           &       7.38 &$\pm$   5.2 & $  -0.4$ --    1.0 & $  -2.5$ --    2.3\\  
 \hline
 \end{tabular}

\caption{\small For each observable listed in the first
column, the second column shows the theoretical expectation
of the ratio $R^{\rm th}$ between 8 and 7 TeV, computed
with NNPDF2.1, and
then the relevant systematic
theoretical uncertainties: PDFs, $\alpha_s$ and
scale variation, computed as discussed
in the text. The theory systematics are given as percentage with respect to
the central prediction. In some cases the theory
systematics in the cross section ratios is at the
sub-permille level, this is indicated as 0.0 in the tables.  \label{tab:8_over_7} }
\end{table}
%%%%%%%%%%%

%%%%%%%%%%%
\begin{table}[t]
\centering
\small
 \begin{tabular}{c|c|c|c|c}
 \hline
 Cross Section & $R^{\rm th,nnpdf}$ & $\delta_{\rm PDF}$(\%)  & $\delta_{\alpha_s}$ (\%)&$\delta_{\rm scales}$ (\%)\\
 \hline
 \hline
$t\bar{t}/Z$                                       &       2.61 &$\pm$   1.6 & $  -1.1$ --    1.0 & $  -0.6$ --    1.4\\  
$t\bar{t}$                                         &       5.58 &$\pm$   1.4 & $  -0.7$ --    0.9 & $  -0.5$ --    1.4\\  
$Z$                                                &       2.14 &$\pm$   0.8 & $  -0.1$ --    0.4 & $  -0.3$ --    0.3\\  
$W^+$                                              &       2.01 &$\pm$   0.8 & $  -0.0$ --    0.3 & $  -0.4$ --    0.3\\  
$W^-$                                              &       2.17 &$\pm$   0.8 & $  -0.1$ --    0.3 & $  -0.4$ --    0.2\\  
$W^+/W^-$                                          &       0.93 &$\pm$   0.4 & $   -0.0$ --    0.1 & $  -0.0$ --    0.1\\  
$W/Z$                                              &       0.97 &$\pm$   0.2 & $  -0.1$ --    0.1 & $  -0.0$ --    0.0\\  
$ggH$                                              &       3.26 &$\pm$   0.8 & $  -0.1$ --    0.2 & $  -1.1$ --    1.1\\  
$t\bar{t}(M_{tt}\ge 1~{\rm TeV})$                  &      14.8 &$\pm$   3.3 & $  -1.0$ --    1.2 & $  -2.2$ --    2.6\\  
$t\bar{t}(M_{\rm tt}\ge 2~{\rm TeV})$              &       69.7 &$\pm$   9.6 & $  -0.6$ --    0.6 & $  -2.8$ --    2.0\\  
$\sigma_{\rm jet}(p_{T}\ge 1~{\rm TeV})$           &       34.9 &$\pm$   2.9 & $   -0.0$ --    0.3 & $  -2.0$ --    2.8\\  
$\sigma_{\rm jet}(p_{T}\ge 2~{\rm TeV})$           &     1340 &$\pm$  12 & $  -0.7$ --    1.1 & $  -8.0$ --    6.4\\  
 \hline
 \end{tabular}

\caption{\small Same as Table~\ref{tab:8_over_7} for ratios between 14
  and 7 TeV LHC center of mass energies. \label{tab:14_over_7} }
\end{table}
%%%%%%%

%%%%%%%%%%
\begin{table}[t]
\centering \small  \begin{tabular}{c|c|c|c|c}
 \hline
 Cross Section & $R^{\rm th,nnpdf}$ & $\delta_{\rm PDF}$(\%)  & $\delta_{\alpha_s}$ (\%)&$\delta_{\rm scales}$ (\%)\\
 \hline
 \hline
$t\bar{t}/Z$                                       &       2.12 &$\pm$   1.3 & $  -0.8$ --    0.8 & $  -0.4$ --    1.1\\  
$t\bar{t}$                                         &       3.90 &$\pm$   1.1 & $  -0.5$ --    0.7 & $  -0.4$ --    1.1\\  
$Z$                                                &       1.84 &$\pm$   0.7 & $  -0.1$ --    0.3 & $  -0.3$ --    0.2\\  
$W^+$                                              &       1.75 &$\pm$   0.7 & $  -0.0$ --    0.3 & $  -0.3$ --    0.2\\  
$W^-$                                              &       1.86 &$\pm$   0.6 & $  -0.1$ --    0.3 & $  -0.3$ --    0.1\\  
$W^+/W^-$                                          &       0.94 &$\pm$   0.3 & $   -0.0$ --    0.0 & $  -0.0$ --    0.0\\  
$W/Z$                                              &       0.98 &$\pm$   0.1 & $  -0.1$ --    0.0 & $  -0.0$ --    0.0\\  
$ggH$                                              &       2.56 &$\pm$   0.6 & $  -0.1$ --    0.1 & $  -0.9$ --    1.0\\  
$t\bar{t}(M_{tt}\ge 1~{\rm TeV})$                  &       8.18 &$\pm$   2.5 & $  -1.3$ --    1.1 & $  -1.6$ --    2.1\\  
$t\bar{t}(M_{\rm tt}\ge 2~{\rm TeV})$              &       24.9 &$\pm$   6.3 & $  -0.0$ --    0.3 & $  -3.0$ --    1.1\\  
$\sigma_{\rm jet}(p_{T}\ge 1~{\rm TeV})$           &       15.1 &$\pm$   2.1 & $  -0.4$ --    0.0 & $  -1.9$ --    2.4\\  
$\sigma_{\rm jet}(p_{T}\ge 2~{\rm TeV})$           &      182 &$\pm$   7.7 & $  -0.3$ --    0.2 & $  -5.7$ --    4.0\\  
 \hline
 \end{tabular}

\caption{\small Same as Table~\ref{tab:8_over_7} 
for ratios between 14 and 8 TeV LHC center of mass energies. \label{tab:14_over_8} }
\end{table}
%%%%%%%%%%%%

\begin{figure}[htb]
   \centering
\epsfig{width=0.49\textwidth,figure=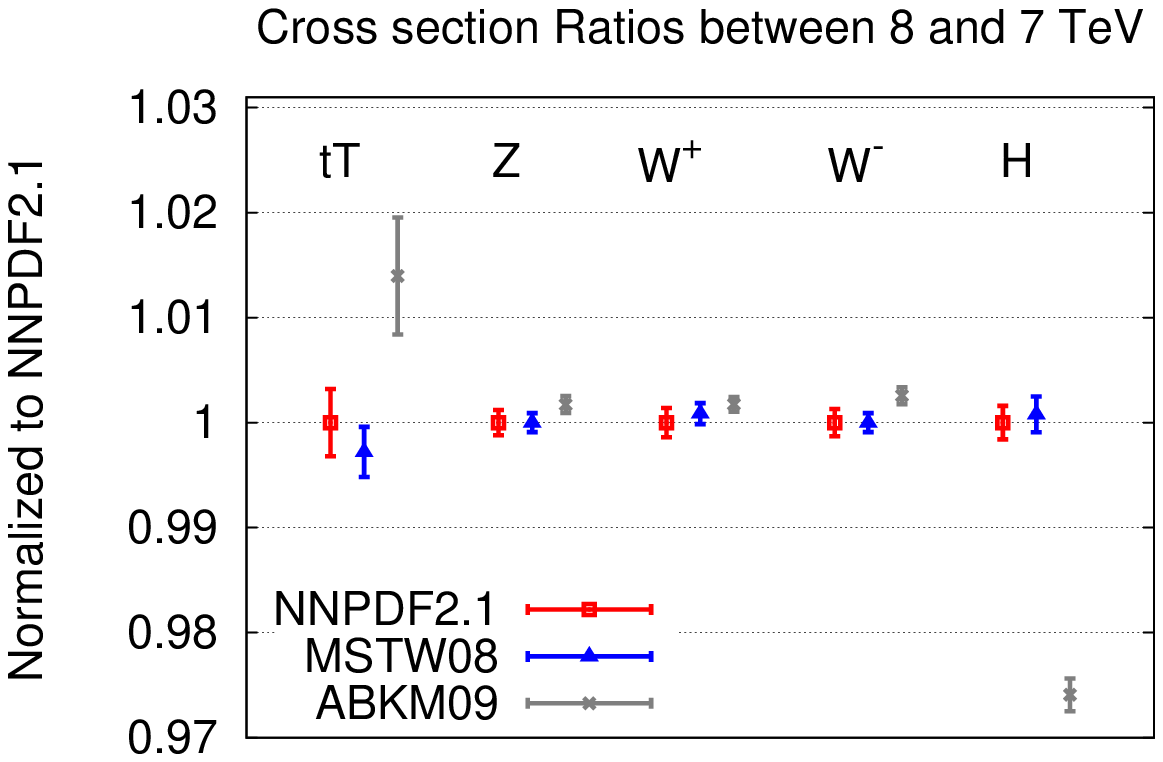}
\epsfig{width=0.49\textwidth,figure=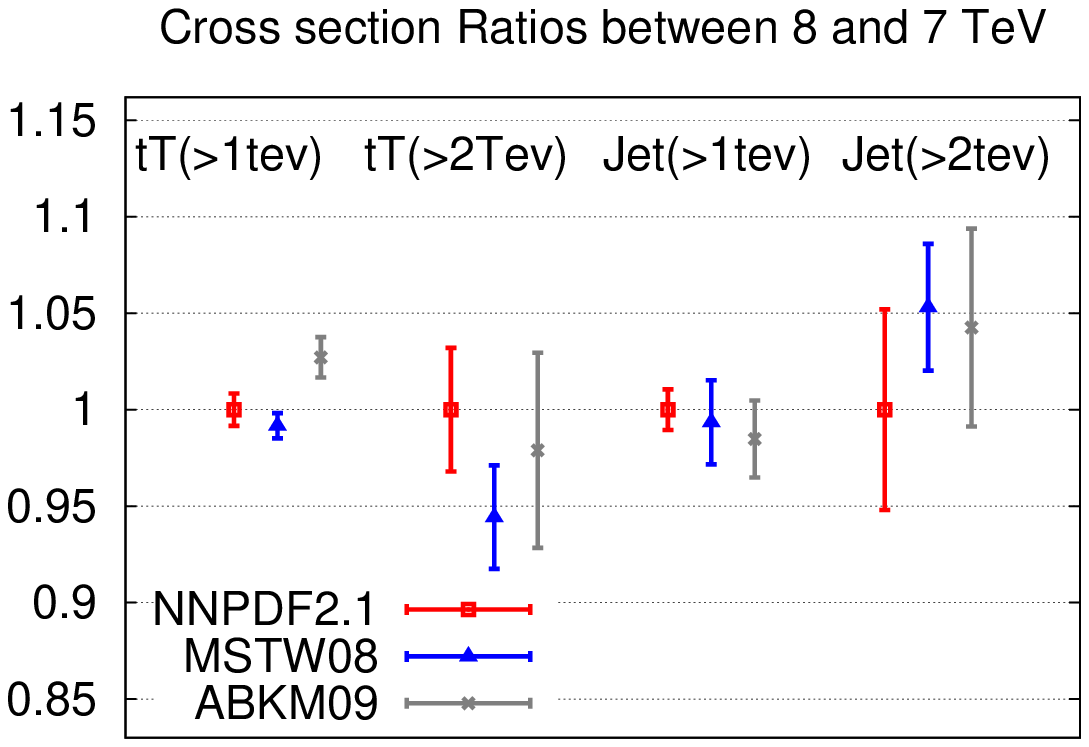}
\epsfig{width=0.49\textwidth,figure=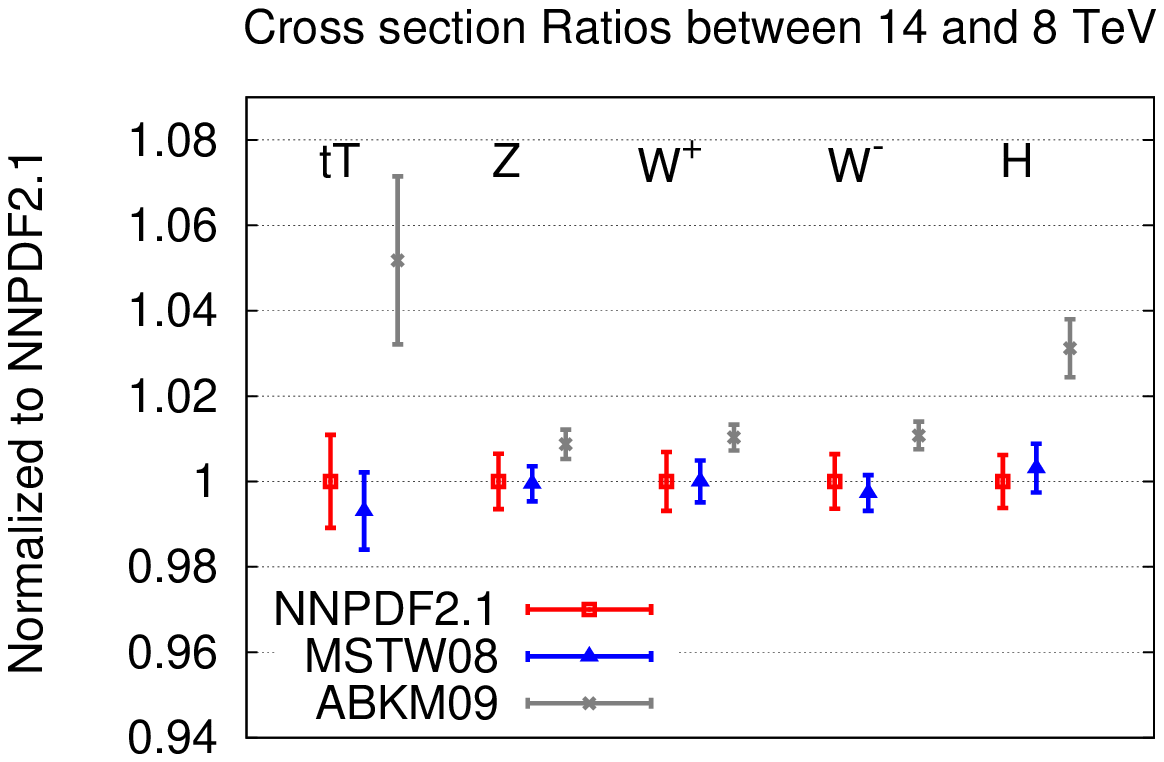}
\epsfig{width=0.49\textwidth,figure=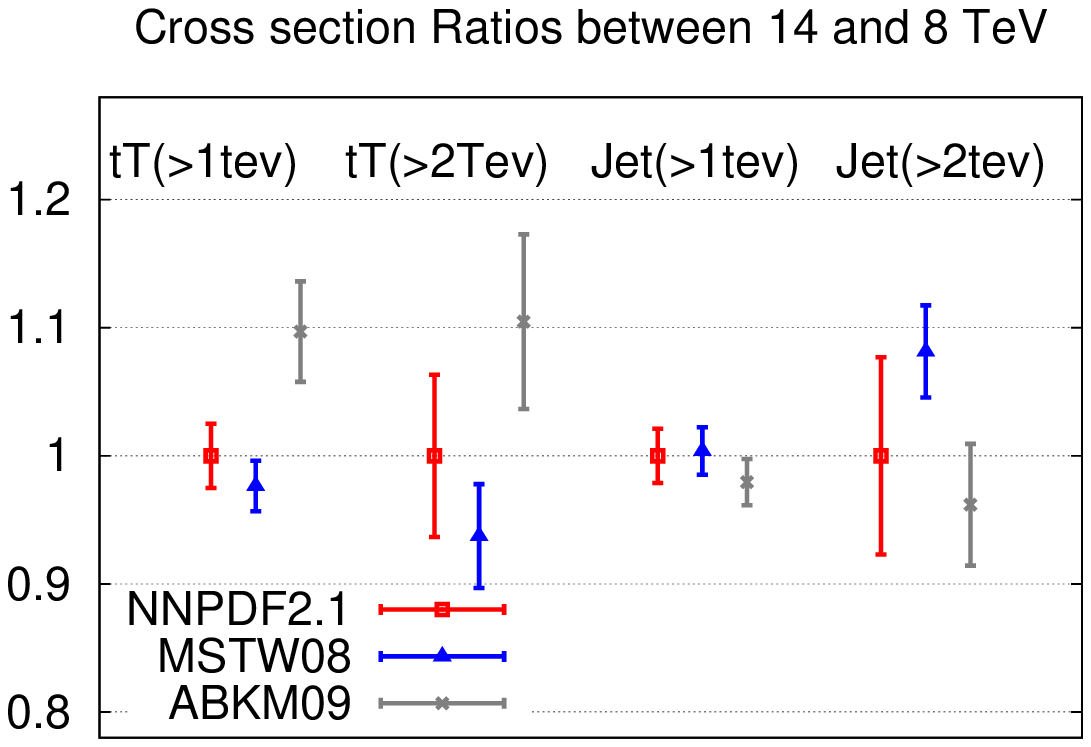}
   \caption{\small Graphical representation of the results
in Tables~\ref{tab:PDF_8_over_7}--\ref{tab:PDF_14_over_8} for
cross section ratios obtained with
different PDF sets. The upper
plots show the results for the cross section ratios of
8 over 7 TeV, obtained
for all three PDF sets considered for the most relevant observables,
normalized to NNPDF2.1 NNLO. The lower plots represent the same
ratios this time for 14 over 8 TeV cross sections. The left plot
show the results for the inclusive cross sections, which
probe $\mathcal{O}\lp 100~{\rm GeV }\rp$ scales, while the 
right plots correspond to more differential distributions
in the $\mathcal{O}\lp 1~{\rm TeV }\rp$ region.}
   \label{fig:ratplot}
\end{figure}

We have collected the results for cross section ratios between 8 and 7
TeV, 14 and 7 TeV, and 14 and 8 TeV, in
Tables~\ref{tab:8_over_7}--\ref{tab:14_over_8}. 
For each process $X$ (or
ratio of processes $X$ and $Y$) we show the theoretical expectation
for $R^{\rm th}(X)$ ($R^{\rm th}(X,Y)$) and the relevant systematic theoretical
uncertainties: PDFs, strong coupling and scales.  
We then studied the stability of our results with
respect to changes in the PDF parameterizations, as shown in Tables~\ref{tab:PDF_8_over_7}--\ref{tab:PDF_14_over_8}, where we collect the
central values, systematics and shifts relative to the reference
NNPDF2.1 NNLO set, obtained by using the MSTW08 and ABKM09 NNLO PDF
sets. Using different PDFs is important to assess the robustness
of the theory prediction, since in some cases differences among PDF 
sets differ by a larger amount than the nominal PDF uncertainty of each set.
The results for a representative subset of
these cross section ratios obtained
with different PDF sets are also represented graphically in
Fig.~\ref{fig:ratplot}.

\begin{figure}[htb]
   \centering
\epsfig{width=0.60\textwidth,figure=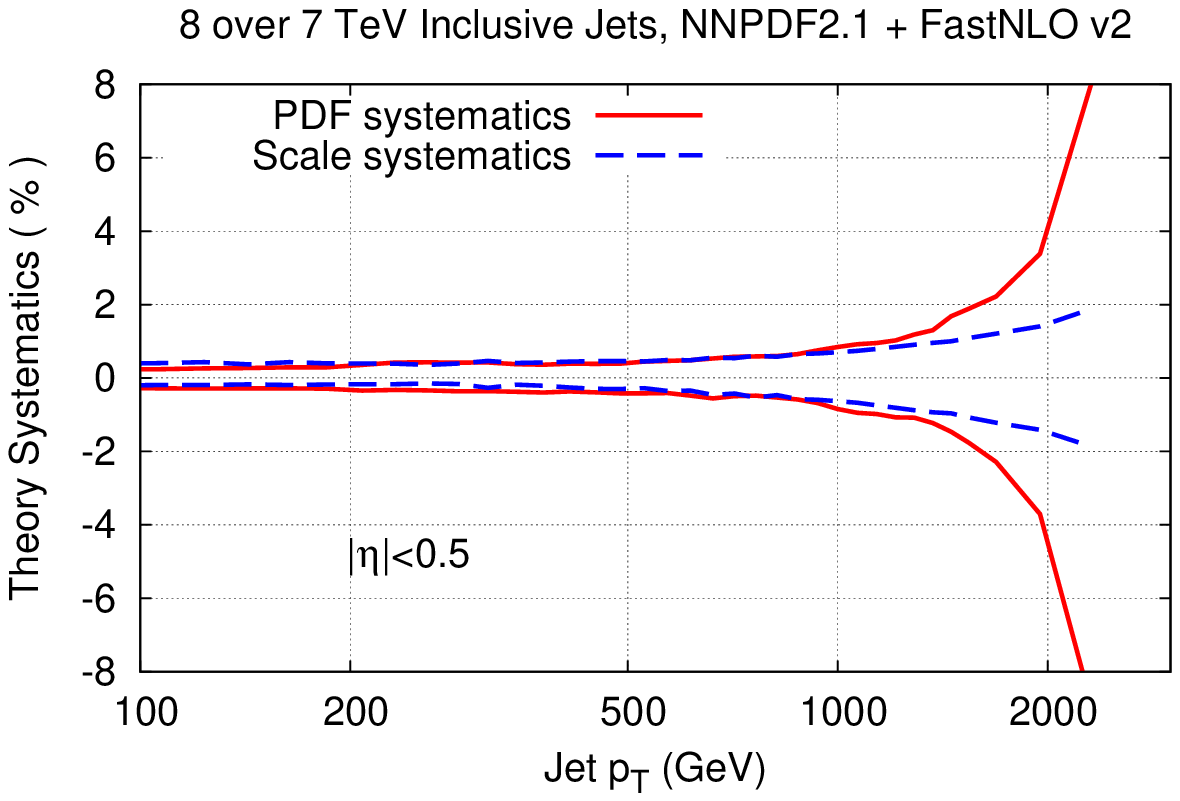}
   \caption{ \small Theory systematics in the 8 over 7 TeV cross section ratios
using {\tt FastNLO}. We show 
the PDF and scale systematics
for the ratio of  8 over 7 TeV cross sections for LHC inclusive jet production
as a function of the $p_T$ of the jet,
 in the central region  $|\eta|\le 0.5$.  }
   \label{fig:DRfastnlo}
\end{figure}

We summarize here the main features of these results. Let us focus
first on the results for $R_{8/7}$:
\begin{itemize}
\item For $W$ and $Z$ production processes, all sources of
  uncertainties have a comparable size, typically of ${\cal
    O}(10^{-3})$ or below. The $W^-$ ratios obtained with the ABKM09
  set differ from NNPDF2.1 and MSTW08 by $2.3\times 10^{-3}$, which is
  nominally a difference of about $2\sigma$, given the individual
  values of $\delta_{\rm PDF}$. This difference however is much
  reduced when considering double ratios ($W^+/W^-$ and $Z/W$), and
  therefore it is unlikely to be measurable, given the uncertainty on
  the relative luminosity calibration at the two energies, which is
  $\sim 2\%$. Notice nevertheless that, since the stability of single
  ratios, for all PDF sets, is at the level of $\sim 2\times 10^{-3}$,
  a precise measurement of $R(Z)$ or $R(W)$ can correlate the
  luminosities of runs at the two energies with this level of accuracy.
\item For inclusive $\ttbar$ production, and for both $R(\ttbar)$ and
  $R(\ttbar,Z)$, $\delta_{\rm scale} \oplus \delta_{\alpha_s} \sim
  4\times 10^{-3}$. The difference between 
  NNPDF2.1 and MSTW08, as well as the individual $\delta_{\rm PDF}$, are of
  similar size, while a shift slightly larger than 1\% is observed
  with respect to ABKM09. This corresponds to a $\sim2.5\sigma$ change,
  thus a potential probe of the gluon PDF parameterizations.
\item For $\ttbar$ production at large mass, $\delta_{\rm scale} \sim
  1\%$, while $\delta_{\rm PDF}$ is of the order of several
  \%, consistent with the intrinsic differences among the different
  PDF sets. $R(\ttbar)$ provides therefore a useful constraint for the gluon
  density at large $x$ (see the discussion of the initial-state
  composition in $\ttbar$ events, in Sect.~\ref{sec:bsm}, where we show
  that high mass $\ttbar$ production is dominated by the $gg$ process).
\item In the case of the jet rates, the scale uncertainty is
  comparable to the PDF one for $p_T>1$~TeV, while the PDF uncertainty
  dominates when $p_T>2$~TeV. This suggest that ratios of high-$p_T$
jet cross sections could be useful to constrain large-$x$ quark PDFs.
To study this possibility in more detail, we have cross-checked
the jet theory systematics in the 8 over 7 TeV cross section ratios
using {\tt FastNLO} with a finer binning of $p_T$ and
rapidity. In Fig.~\ref{fig:DRfastnlo} we show 
the PDF and scale systematics for LHC inclusive jet production
as a function of the $p_T$ of the jet, in the central region  $|\eta|\le 0.5$.
As can be seen, PDF and scale systematics are below 1\% below 1 TeV,
and while scale systematics are small even for larger $p_T$, at some
point near $p_T \sim 2.5$ TeV (corresponding to a final state
with approximately $m_X\sim 5$ TeV in the central
region) the PDF uncertainties blow up:
therefore, measurements in this region would be important
to constrain large-$x$ PDFs.
\end{itemize}

Considering the ratios at 14 and 8 TeV, the following additional
remarks can be made:
\begin{itemize}
\item For electroweak processes, all uncertainties grow slightly, but
  still remain well below 1\% in the case of NNPDF2.1 and
  MSTW08. Rate ratios obtained with ABKM09 are about 1\% smaller,
  which is a $\sim 2\sigma$ effect. Once again, the
  measurement of these ratios provides a very effective tool to calibrate
  at the percent level the relative normalization of the 8~TeV and 14~TeV
  absolute luminosities.  
\item For $R(\ttbar)$, the scale systematics is $\sim 1\%$. As in the
  case of the 8/7 ratios, PDF
  differences between NNPDF2.1 and MSTW08 are compatible with the
  individual values of $\delta_{\rm PDF}$, which are also $\sim
  1\%$. The value of $R(\ttbar)$ obtained with ABKM09 is $\sim 5\%$
  smaller, corresponding to a $\sim 2.5\sigma$ effect.
\item For $\ttbar$ production at large mass, $\delta_{\rm scale} \sim
  2-3\%$, while $\delta_{\rm PDF}$ grows to $6\%$, showing a great
  sensitivity to the PDF distributions.
\item The gluon fusion Higgs production cross sections has very small
  PDF and scale systematics in the 14 over 8 TeV ratio. Therefore,
  measurements of this ratio could provide stringent tests of the
  hypothesis that the measured Higgs boson indeed behaves as a
  Standard Model Higgs boson from the production point of view. As
  suggested in~\cite{Krasny:2011dj}, consideration of rate ratios for
  individual Higgs decay final states (e.g. $WW^* \to 2\ell 2\nu$)
  could also be used to consolidate the separation of signal and
  backgrounds, due to the different energy scaling of the respective rates.
\item Scale and PDF uncertainties are comparable in the case of jet
  production, for both thresholds of $p_T>1$~TeV and $p_T>2$~TeV. This
  is the result of the rather different composition of the initial
  state at the two energies (see Sect.~\ref{sec:bsm}), 
so that the scale dependence at the two energies is only weakly correlated.
\end{itemize}

%%%%%%%%%
\begin{table}[t]
\centering
\scriptsize
 \begin{tabular}{c|c|c|c|c|c|c|c|c}
 \hline
 Ratio & $R^{\rm nnpdf}$ & $\delta_{\rm PDF}$(\%)  & $R^{\rm mstw}$ & $\delta_{\rm PDF}$(\%)  & $\Delta^{\rm mstw} (\%) $ & $R^{\rm abkm}$ & $\delta_{\rm abkm}$(\%)  & $\Delta^{\rm abkm}$ (\%) \\ 
 \hline
 \hline
$t\bar{t}/Z$                                       &       1.23 &   0.4 &       1.23 &   0.2 &   $+$0.3 &       1.25 &   0.5 &  $-$1.3\\  
$t\bar{t}$                                         &       1.43 &   0.3 &       1.43 &   0.2 &   $+$0.3 &       1.45 &   0.5 &  $-$1.4\\  
$Z$                                                &       1.16 &   0.1 &       1.16 &   0.1 &  $+$0.0 &       1.16 &   0.1 &  $-$0.1\\  
$W^+$                                              &       1.15 &   0.1 &       1.15 &   0.1 &  $-$0.1 &       1.15 &   0.1 &  $-$0.2\\  
$W^-$                                              &       1.17 &   0.1 &       1.17 &   0.1 &   $+$0.0 &       1.17 &   0.1 &  $-$0.2\\  
$W^+/W^-$                                          &       0.98 &   0.1 &       0.98 &   0.0 &  $-$0.1 &       0.98 &   0.0 &   $+$0.0\\  
$W/Z$                                              &       0.99 &   0.0 &       0.99 &   0.0 &  $-$0.0 &       0.99 &   0.0 &   $+$0.0\\  
$ggH$                                              &       1.27 &   0.2 &       1.27 &   0.2 &  $-$0.1 &       1.24 &   0.2 &   $+$2.6\\  
$t\bar{t}(M_{tt}\ge 1~{\rm TeV})$                  &       1.81 &   0.8 &       1.79 &   0.7 &   $+$0.9 &       1.86 &   1.0 &  $-$2.7\\  
$t\bar{t}(M_{\rm tt}\ge 2~{\rm TeV})$              &       2.80 &   3.2 &       2.64 &   2.8 &   $+$5.7 &       2.74 &   5.2 &   $+$2.3\\  
$\sigma_{\rm jet}(p_{T}\ge 1~{\rm TeV})$           &       2.30 &   1.0 &       2.29 &   2.2 &   $+$0.3 &       2.27 &   2.0 &   $+$1.1\\  
$\sigma_{\rm jet}(p_{T}\ge 2~{\rm TeV})$           &       7.38 &   5.2 &       7.77 &   3.1 &  $-$4.5 &       7.69 &   4.9 &  $-$3.5\\  
 \hline
 \end{tabular}

\caption{\small For each observable listed in the first column, we
  show the theoretical predictions for the various observables ratios
  between 8 and 7 TeV when computing with the NNPDF2.1, MSTW08 and
  ABKM09 NNLO PDF sets, as well as the respective percentage PDF error
$\delta_{\rm PDF}$ in each
  case and the percentage shift with respect the NNPDF2.1 prediction.
 When the shift with respect to NNPDF2.1 is below the permille 
level it is denoted by 0.0 in the table.
  In each case the default value of $\alpha_s(M_Z)$ provided by MSTW08
  ($\alpha_s(M_Z)=0.1171$) and ABKM09 ($\alpha_s(M_Z)=0.1135$) have been used.
to be compared with $\alpha_s(M_Z)=0.119$ in the baseline NNPDF2.1
predictions.
 \label{tab:PDF_8_over_7} }
\end{table}
%%%%%%%%

%%%%%%%%%%%%%%%%%%%%%%%%%%%%%%%%%%%%%%%%%%
\begin{table}[t]
\centering
\scriptsize
 \begin{tabular}{c|c|c|c|c|c|c|c|c}
 \hline
 Ratio & $R^{\rm nnpdf}$ & $\delta_{\rm PDF}$(\%)  & $R^{\rm mstw}$ & $\delta_{\rm PDF}$(\%)  & $\Delta^{\rm mstw} (\%) $ & $R^{\rm abkm}$ & $\delta_{\rm abkm}$(\%)  & $\Delta^{\rm abkm}$ (\%) \\ 
 \hline
 \hline
$t\bar{t}/Z$                                       &       2.61 &   1.6 &       2.59 &   1.2 &   $+$0.9 &       2.76 &   2.4 &  $-$5.6\\  
$t\bar{t}$                                         &       5.58 &   1.4 &       5.53 &   1.2 &   $+$1.0 &       5.96 &   2.4 &  $-$6.7\\  
$Z$                                                &       2.14 &   0.8 &       2.14 &   0.5 &   $+$0.1 &       2.16 &   0.4 &  $-$1.0\\  
$W^+$                                              &       2.01 &   0.8 &       2.01 &   0.6 &  $-$0.1 &       2.03 &   0.4 &  $-$1.2\\  
$W^-$                                              &       2.17 &   0.8 &       2.16 &   0.5 &   $+$0.3 &       2.20 &   0.4 &  $-$1.3\\  
$W^+/W^-$                                          &       0.93 &   0.4 &       0.93 &   0.2 &  $-$0.4 &       0.92 &   0.2 &   $+$0.1\\  
$W/Z$                                              &       0.97 &   0.2 &       0.97 &   0.1 &  $-$0.0 &       0.97 &   0.1 &  $-$0.2\\  
$ggH$                                              &       3.26 &   0.8 &       3.28 &   0.7 &  $-$0.4 &       3.28 &   0.8 &  $-$0.4\\  
$t\bar{t}(M_{tt}\ge 1~{\rm TeV})$                  &      14.8 &   3.3 &      14.3 &   2.3 &   $+$3.3 &      16.6 &   4.1 & $-$12.5\\  
$t\bar{t}(M_{\rm tt}\ge 2~{\rm TeV})$              &       69.7 &   9.6 &       61.7 &   6.0 &  $+$11.9 &       75.4 &  10.1 &  $-$7.6\\  
$\sigma_{\rm jet}(p_{T}\ge 1~{\rm TeV})$           &       34.9 &   2.9 &       34.8 &   2.1 &  $-$0.2 &       33.6 &   2.2 &   $+$3.1\\  
$\sigma_{\rm jet}(p_{T}\ge 2~{\rm TeV})$           &     1340 &  12.4 &     1527 &   4.0 & $-$11.5 &     1344 &   6.2 &   $+$1.8\\  
 \hline
 \end{tabular}

\caption{\small Same as Table~\ref{tab:PDF_8_over_7} 
but for cross section ratios between 14 and 7 TeV.
\label{tab:PDF_14_over_7} }
\end{table}
%%%%%%%%%%%%%%%%%%%%%%%%%%%%%%%%%%%%%%%%%%

%%%%%%%%%%%%%%%%%%%%%%%%%%%%%
\begin{table}[t]
\centering
\scriptsize
 \begin{tabular}{c|c|c|c|c|c|c|c|c}
 \hline
 Ratio & $R^{\rm nnpdf}$ & $\delta_{\rm PDF}$(\%)  & $R^{\rm mstw}$ & $\delta_{\rm PDF}$(\%)  & $\Delta^{\rm mstw} (\%) $ & $R^{\rm abkm}$ & $\delta_{\rm abkm}$(\%)  & $\Delta^{\rm abkm}$ (\%) \\ 
 \hline
 \hline
$t\bar{t}/Z$                                       &       2.12 &   1.3 &       2.11 &   0.9 &   $+$0.6 &       2.21 &   1.9 &  $-$4.3\\  
$t\bar{t}$                                         &       3.90 &   1.1 &       3.87 &   0.9 &   $+$0.7 &       4.10 &   1.9 &  $-$5.2\\  
$Z$                                                &       1.84 &   0.7 &       1.84 &   0.4 &   $+$0.1 &       1.85 &   0.3 &  $-$0.8\\  
$W^+$                                              &       1.75 &   0.7 &       1.75 &   0.5 &  $+$0.0 &       1.77 &   0.3 &  $-$1.0\\  
$W^-$                                              &       1.86 &   0.6 &       1.85 &   0.4 &   $+$0.3 &       1.88 &   0.3 &  $-$1.1\\  
$W^+/W^-$                                          &       0.94 &   0.3 &       0.94 &   0.2 &  $-$0.3 &       0.94 &   0.1 &   $+$0.0\\  
$W/Z$                                              &       0.98 &   0.1 &       0.98 &   0.1 &   $+$0.0 &       0.98 &   0.1 &  $-$0.2\\  
$ggH$                                              &       2.56 &   0.6 &       2.57 &   0.6 &  $+$0.3 &       2.64 &   0.7 &  $-$3.1\\  
$t\bar{t}(M_{tt}\ge 1~{\rm TeV})$                  &       8.18 &   2.5 &       7.99 &   2.0 &   $+$2.5 &       8.97 &   3.6 &  $-$9.6\\  
$t\bar{t}(M_{\rm tt}\ge 2~{\rm TeV})$              &       24.9 &   6.3 &       23.3 &   4.3 &   $+$6.4 &       27.5 &   6.2 & $-$10.3\\  
$\sigma_{\rm jet}(p_{T}\ge 1~{\rm TeV})$           &       15.1 &   2.1 &       15.2 &   1.9 &  $-$0.5 &       14.8 &   1.8 &   $+$1.9\\  
$\sigma_{\rm jet}(p_{T}\ge 2~{\rm TeV})$           &      181.6 &   7.7 &      196.4 &   3.3 &  $-$7.1 &      174.7 &   4.9 &   $+$4.7\\  
 \hline
 \end{tabular}

\caption{\small Same as Table~\ref{tab:PDF_8_over_7} 
but for cross section ratios between 14 and 8 TeV.
\label{tab:PDF_14_over_8}  }
\end{table}
%%%%%%%%%%%%%%%%%%%%%%%%%%%%%

%%%%%%%%%%%%%%%%%%%%%%%
%%%%%%%%%%%%%%%%%%%%

%%%%%%%%%%%%%%%%%%%%%%%
%%%%%%%%%%%%%%%%%%%%

\section{Parton luminosity ratios}
\label{sec:lumi}

In order to understand better the behavior of cross section
ratios and their PDF systematics presented in the previous section,
as well as to maximize the sensitivity to BSM effects, it
is useful to consider parton--parton luminosities~\cite{Campbell:2006wx}.
Parton luminosities encode the essential information from the
partonic contribution for different subprocesses.
We define four different parton luminosities:\footnote{One can define
other partonic luminosities for more specific processes, like the
$bg$ luminosity, but the four that we discuss are enough for the
most important processes.}

\begin{itemize}

\item Gluon-Gluon luminosity:
\be
\label{eq:lumi1}
\mathcal{L}_{gg}\lp M,s \rp \equiv \frac{1}{s}\int_{\tau}^1
\frac{dx}{x} g\lp x,M\rp g\lp \tau/x,M\rp
\ee

\item Quark-Gluon luminosity
\bea
\label{eq:lumi2}
\mathcal{L}_{gq}\lp M,s \rp &\equiv& \frac{1}{s}\int_{\tau}^1
\frac{dx}{x} \sum_{i=1}^{n_f} \Bigg[ g\lp x,M\rp \lp q_i\lp \tau/x,M\rp
+\bar{q}_i\lp \tau/x,M\rp\rp  \nonumber \\ 
&+&\lp q_i\lp x,M\rp
+\bar{q}_i\lp x,M\rp\rp g\lp \tau/x,M\rp \Bigg]  
\eea

\item Quark-Antiquark luminosity
\be
\label{eq:lumi3}
\mathcal{L}_{q\bar{q}}\lp M,s \rp \equiv \frac{1}{s}\int_{\tau}^1
\frac{dx}{x} \lc \sum_{i,j=1}^{n_f} \lp q_i\lp x,M\rp\bar{q}_j\lp \tau/x,M\rp
+ \bar{q}_i\lp x,M\rp q_j\lp \tau/x,M\rp\rp \rc
\ee

\item Quark-Quark luminosity
\be
\label{eq:lumi4}
\mathcal{L}_{qq}\lp M,s \rp \equiv \frac{1}{s}\int_{\tau}^1
\frac{dx}{x} \lc \sum_{i,j=1}^{n_f} q_i\lp x,M\rp q_j\lp \tau/x,M\rp  \rc
\ee

\end{itemize}
In the above definitions, $\tau=M^2/s$, $M$ is the invariant mass of the produced
final state and $\sqrt{s}$ is the hadronic center of mass energy. Also
$n_f$ is the number of active quark flavors at scale $M$.  
This definition includes the contribution of the top quark PDFs,
but we have verified that PDF luminosities defined for a $n_f=5$
scheme (where top is always considered a massive parton) are
very similar in the relevant kinematical region.

Of particular interest in our case are the ratios of parton luminosities
between different LHC center of mass energies. For the time being
we concentrate on 14 over 8 TeV ratios and 8 over 7 TeV ratios, the
most relevant ones from the phenomenological point of view.
These PDF luminosity ratios are defined as:

\begin{itemize}

\item Gluon-Gluon luminosity ratio
\be
\label{eq:rlumi1}
\mathcal{R}_{gg}\lp M,s_2,s_1 \rp \equiv \mathcal{L}_{gg}\lp M,s_2 \rp /
\mathcal{L}_{gg}\lp M,s_1 \rp 
\ee
\item Quark-Gluon luminosity ratio
\be
\label{eq:rlumi2}
\mathcal{R}_{gq}\lp M,s_2,s_1 \rp \equiv \mathcal{L}_{gq}\lp M,s_2 \rp /
\mathcal{L}_{gq}\lp M,s_1 \rp 
\ee
\item Quark-Antiquark luminosity ratio
\be
\label{eq:rlumi3}
\mathcal{R}_{q\bar{q}}\lp M,s_2,s_1 \rp \equiv \mathcal{L}_{q\bar{q}}\lp M,s_2 \rp /
\mathcal{L}_{q\bar{q}}\lp M,s_1 \rp 
\ee
\item Quark-Quark luminosity ratio
\be
\label{eq:rlumi4}
\mathcal{R}_{qq}\lp M,s_2,s_1 \rp \equiv \mathcal{L}_{qq}\lp M,s_2 \rp /
\mathcal{L}_{qq}\lp M,s_1 \rp 
\ee
\end{itemize}

\begin{figure}[htb]
   \centering
\epsfig{width=0.49\textwidth,figure=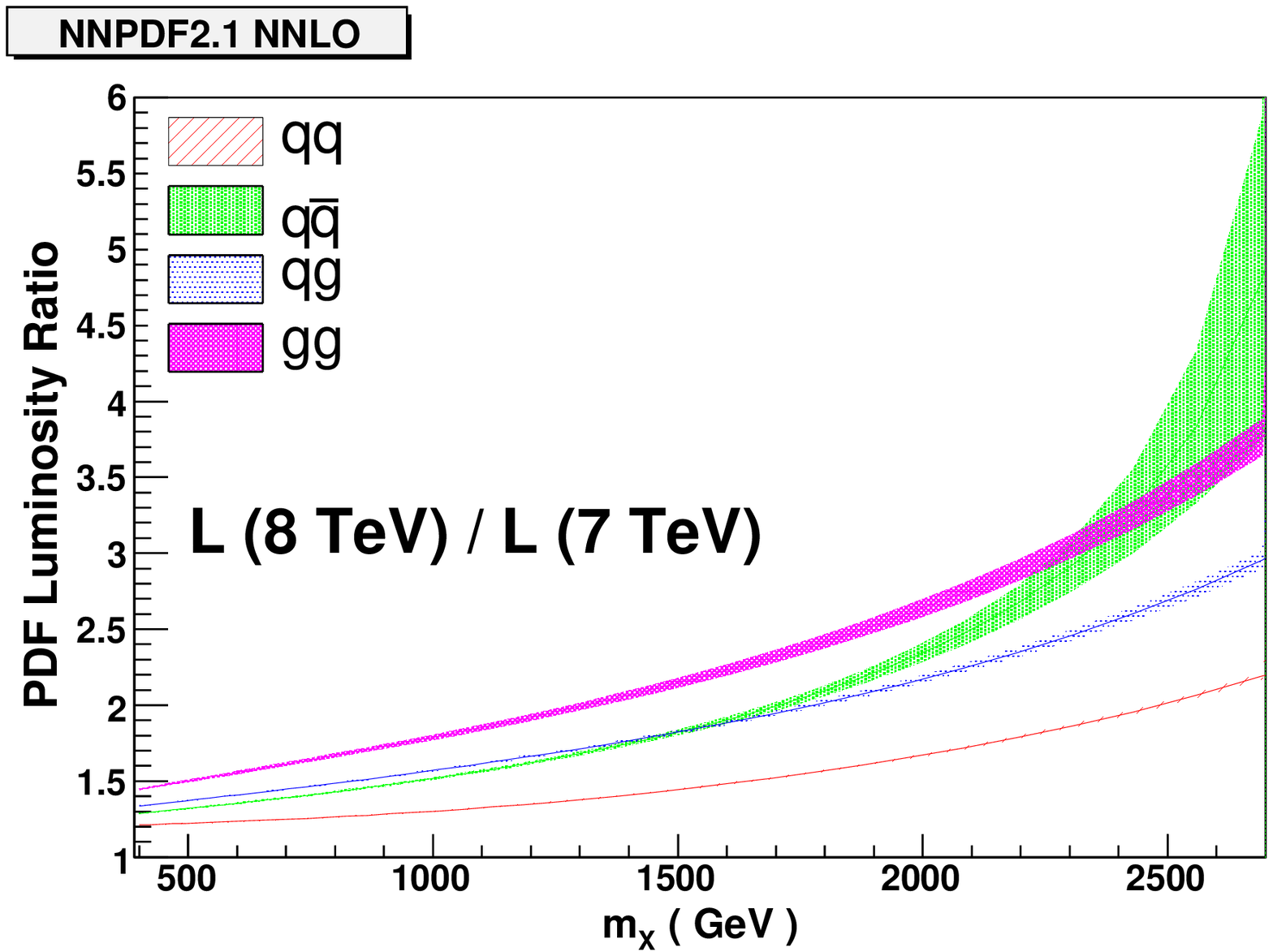}
\epsfig{width=0.49\textwidth,figure=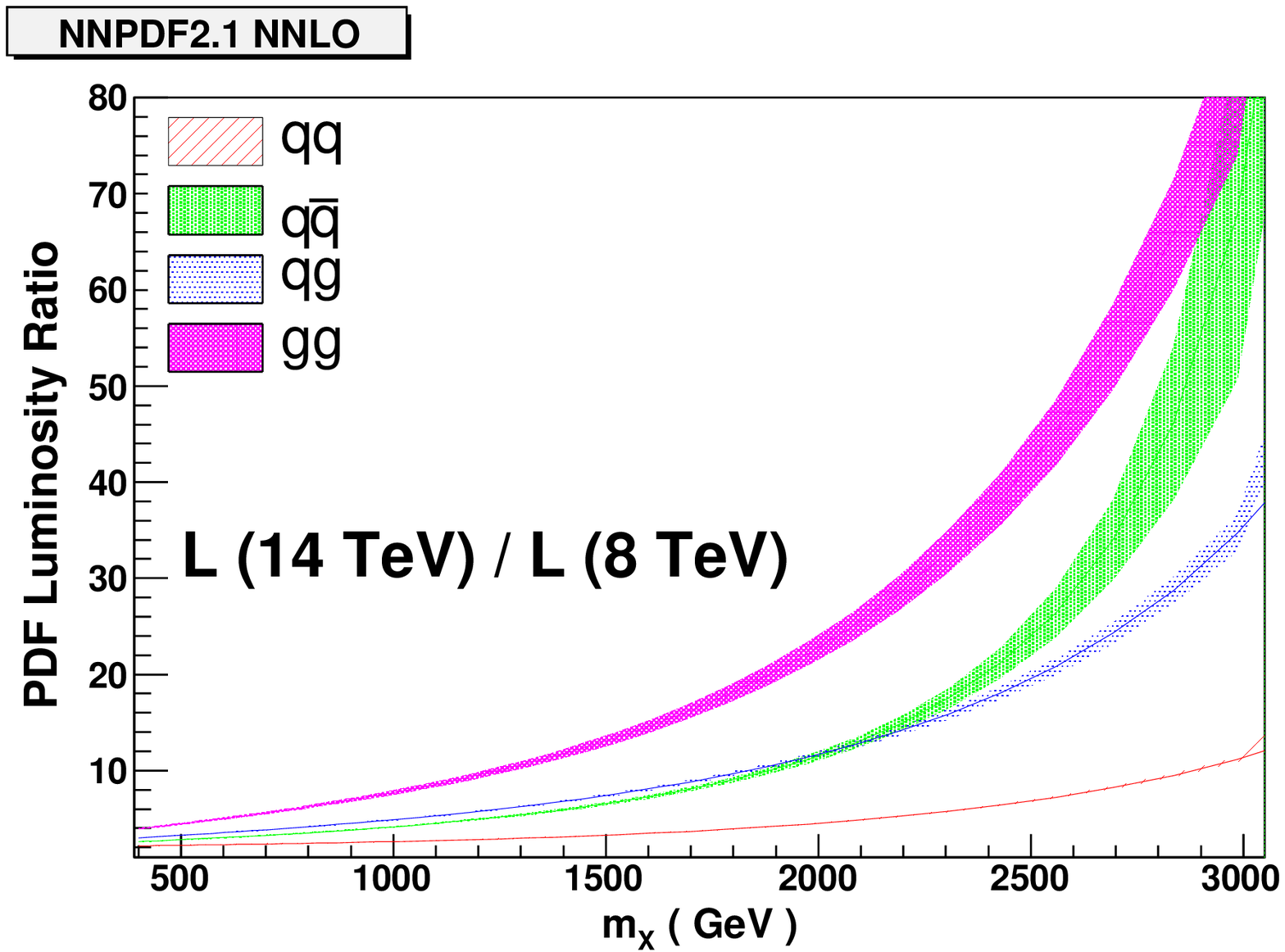}
\epsfig{width=0.49\textwidth,figure=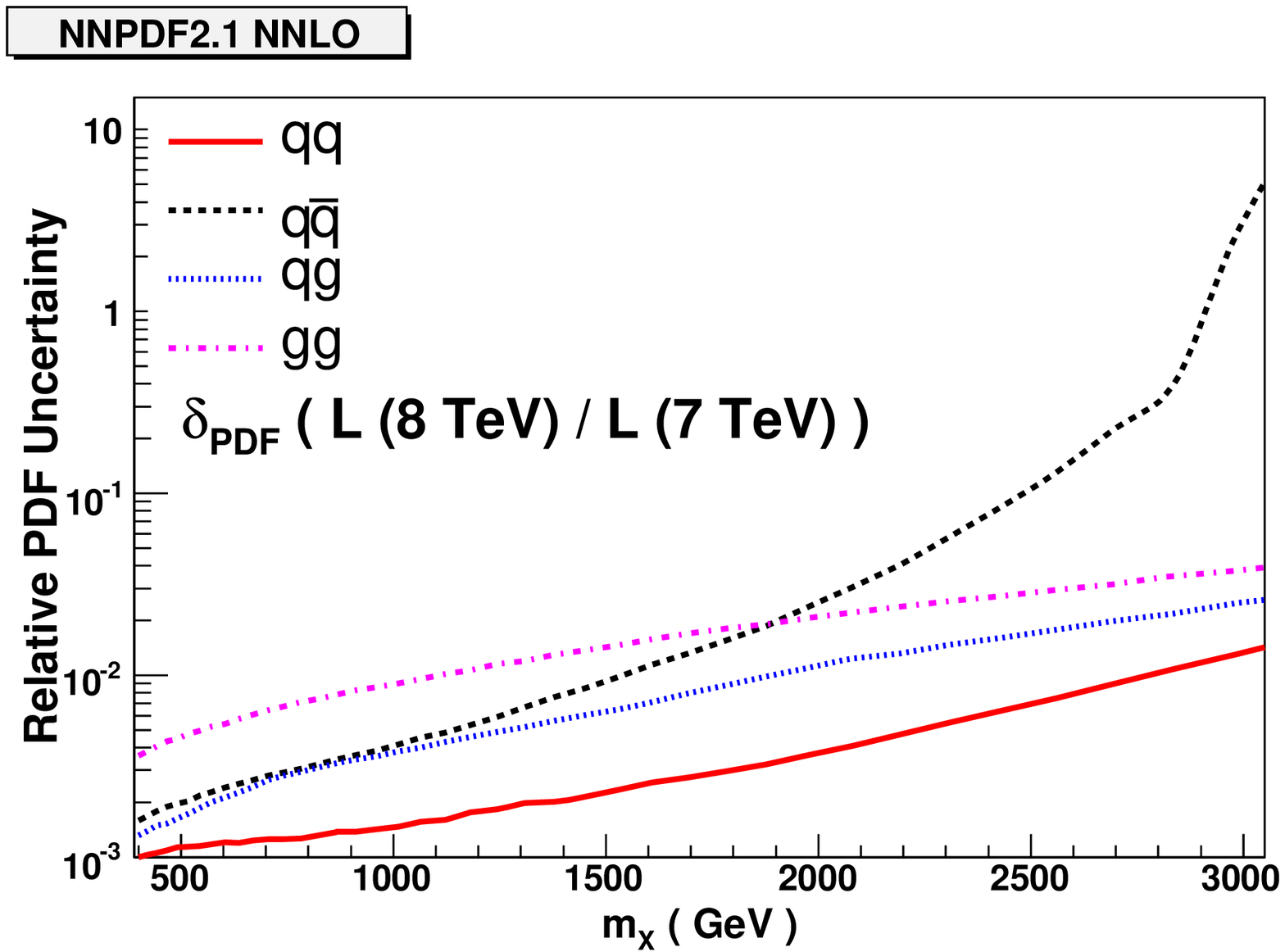}
\epsfig{width=0.49\textwidth,figure=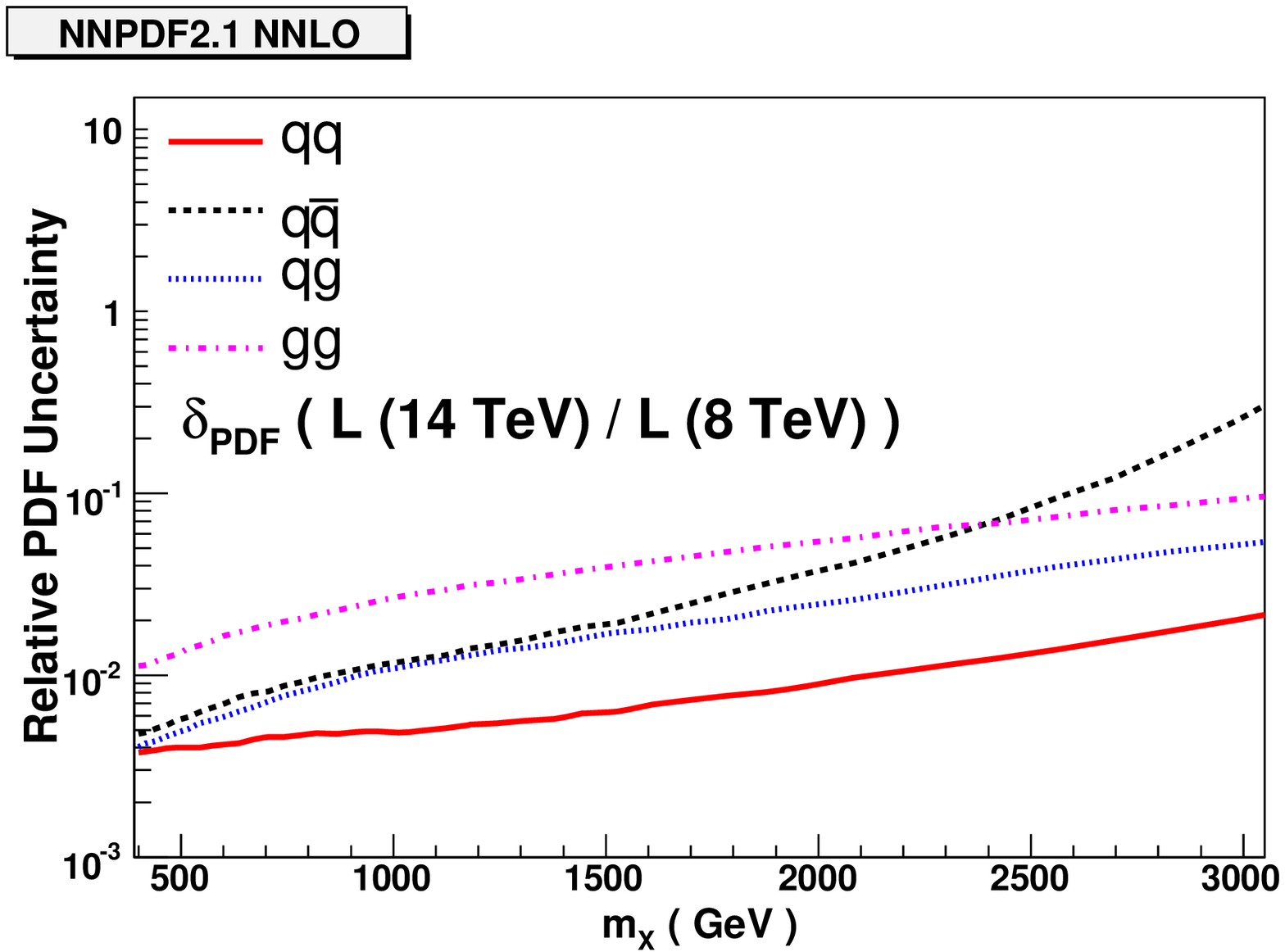}
   \caption{ \small Upper plots: Ratios of PDF luminosities, 
Eqns.~(\ref{eq:rlumi1}--\ref{eq:rlumi4}), between different LHC
beam energies. Lower plots: the relative PDF error $\delta_{\rm PDF}$ 
for each of the above
luminosity ratios. The left plots show the 8 over 7 TeV ratios while
the right plots correspond to the 14 over 8 ratios.}
   \label{fig:lumirat}
\end{figure}

In Fig.~\ref{fig:lumirat} we show the PDF luminosity ratios, defined as above, for the 8 over 7 TeV
ratios and for the 14 over 8 TeV ratios. They have been obtained with NNPDF2.1 NNLO, and the
PDF uncertainties have been obtained from the 1000-replica set. PDF errors
are computed using 68\% Confidence Level intervals to avoid possible
non-gaussian behaviors for large final state masses. We also plot
in Fig.~\ref{fig:lumirat} the percentage PDF errors on these luminosity ratios.

%%%%%%%%%%%%%%%%%%%%%%%
%%%%%%%%%%%%%%%%%%%%

From Fig.~\ref{fig:lumirat} we can see that, as well known, the ratio
of luminosities increases when the beam energy is increased,
growing with the mass of the final state produced particles.
This enhancement is a factor between 1.5 and 5 for final
states with $m_X$ between 0.5 and 3 TeV, depending on the dominant
partonic subprocesses, and a factor between 2 and 80 in the same
range for ratios of 14 over 8 TeV. What is perhaps not so well
appreciated is that PDF uncertainties cancel to a very good
extent in the ratio, for example, for $m_X$ below 1.5 TeV
the PDF uncertainties in the 8 over 7 TeV luminosity ratio
are well below the percent level, confirming the findings
of Tables~\ref{tab:8_over_7}--\ref{tab:14_over_8}.

On the other hand, for large invariant masses the cancellation
of PDF uncertainties breaks down and PDF errors can become
much larger. For the 8 over 7 TeV ratio, for example, the
$q\bar{q}$ luminosity has a very large PDF error, larger
than 100\%, above $m_X=3$ TeV. This is so because in this region
one is probing the antiquark PDFs at very large $x$, a region
in which these PDFs are virtually unknown. Is clear thus that
the measurement of cross section ratios that involve high mass final
states provides stringent constraints on large--$x$ PDFs, which
in turn are an important ingredient for new physics
searches like supersymmetric particle production~\cite{susy}.

Let us conclude this section by mentioning that the qualitative
behavior of the parton luminosity ratios is very similar
if the MSTW08 PDF set is used instead.

\section{Sensitivity to BSM contributions}
\label{sec:bsm}
Having evaluated the systematic uncertainties of the
cross section ratios of relevant LHC cross sections, and having
seen that they are very small in general, we would like to discuss
how the study of these ratios could allow to detect 
possible Beyond the Standard Model (BSM) contributions, 
that might be not accessible through absolute cross sections.

If the final state $X$ receives
contributions from both SM and BSM processes, we shall write:
\be
\sigma(pp\to X) = \sigma^{SM}(pp\to X) + \sigma^{\rm BSM}(pp\to X) \; ,
\ee
and, under the assumption that the BSM contamination represents only a
small fraction of the total, 
\be
\label{eq:rmaster}
R^X_{E_1/E_2} \sim  \frac{\sigma^{\rm SM}_X(E_1)}{\sigma^{\rm SM}_X(E_2)} \times \left\{1+ 
\frac{\sigma^{\rm BSM}_X(E_1)}{\sigma^{\rm SM}_X(E_1)} \; \Delta_{E_1/E_2} \left[
\frac{\sigma^{\rm BSM}_X}{\sigma^{\rm SM}_X}  \right] \right\}  \; ,
\ee
where we defined, for a quantity $A$:
\be
\Delta_{E_1/E_2} (A) =1-\frac{A(E_2)}{A(E_1)} \; .
\ee
The above equations translate in formulas the obvious observation
that the visibility of a BSM contribution in the evolution with energy
of $\sigma(X)$ requires that it evolves with energy differently than the SM
one: if $\sigma^{\rm BSM}(pp\to X)/\sigma^{\rm SM}(pp\to X)$ were independent
of $E$, no information could be obtained from the study of the energy
evolution of $\sigma(X)$. 
The threshold for the visibility of such effects
is given by the precision of the SM prediction, which sets the overall
theoretical systematics, and defines the goals of the
experimental precision:
\be
\frac{\sigma^{\rm BSM}_X(E_1)}{\sigma^{\rm SM}_X(E_1)} \;  \times \; \Delta_{E_1/E_2} \left[
\frac{{\sigma}^{\rm BSM}_X}{{\sigma}^{\rm SM}_X}  \right]  
>
\delta_{TH} \equiv \frac{\delta R^{\rm SM}_{E_1/E_2}}{R^{\rm SM}_{E_1/E_2}} \; .
\ee
Having established in the previous Section that $\delta_{TH}$ is
typically at the percent level, and in some cases
at the permille level, BSM contributions of few $\%$ could be
detected if $\sigma^{\rm BSM}(pp\to X)$ and $\sigma^{\rm SM}(pp\to X)$ have a
sufficiently different energy scaling. 
In addition to the matrix-element structure, which may vary from process
to process, the energy scaling depends on the initial state
partons $(i,j)$, 
which define the partonic luminosity ${\cal L}_{ij}$, 
 Eqns.~(\ref{eq:rlumi1}--\ref{eq:rlumi4}),
as discussed in the previous section.

To give an
example, consider the production of a final state $X$ of mass $M$.
Assuming that
\be
\sigma^{\rm SM}(X) \sim {\cal L}_{ab}(M) \hat{\sigma}^{\rm SM}(X,M) \; ,
\quad
\sigma^{\rm BSM}(X) \sim {\cal L}_{ij}(M) \hat{\sigma}^{\rm BSM}(X,M) \; ,
 \ee
where the partonic cross sections $\hat{\sigma}^{\rm SM,BSM}(X,M)$ depend
on $M$, but are
independent of beam energy, we obtain:
\be
\label{eq:deltamaster}
\Delta_{E_1/E_2} \left[
\frac{{\sigma}^{\rm BSM}_X}{{\sigma}^{\rm SM}_X}  \right]  \sim
\Delta_{E_1/E_2} \left[
\frac{{\cal L}_{ ij}(M)}{{\cal L}_{ ab}(M)}  \right] = 
1-\frac{\mathcal{L}_{ ij}(M,E_2)/\mathcal{L}_{ab}(M,E_2)}{\mathcal{L}_{ij}(M,E_1)/\mathcal{L}_{ ab}(M,E_1)} \; .
\ee
The energy dependence of luminosity ratios could therefore expose the
possible existence of BSM phenomena via the study of cross section
energy ratios.

We illustrate these considerations with three 
 examples: top quark pair production at large
$\ttbar$ pair masses, inclusive jet production at large
transverse momentum, and high-mass off-shell
$Z$-boson production. These processes
are dominated by different production channels, $gg$ for
$\ttbar$, $qq$ for jets and $q\bar{q}$ for 
$Z$ production, and are amongst the cross
sections that have or will be measured with high precision
in the TeV regime, where the sensitivity to new physics
is enhanced.

\begin{figure}[t]
  \centering
\epsfig{width=0.49\textwidth,figure=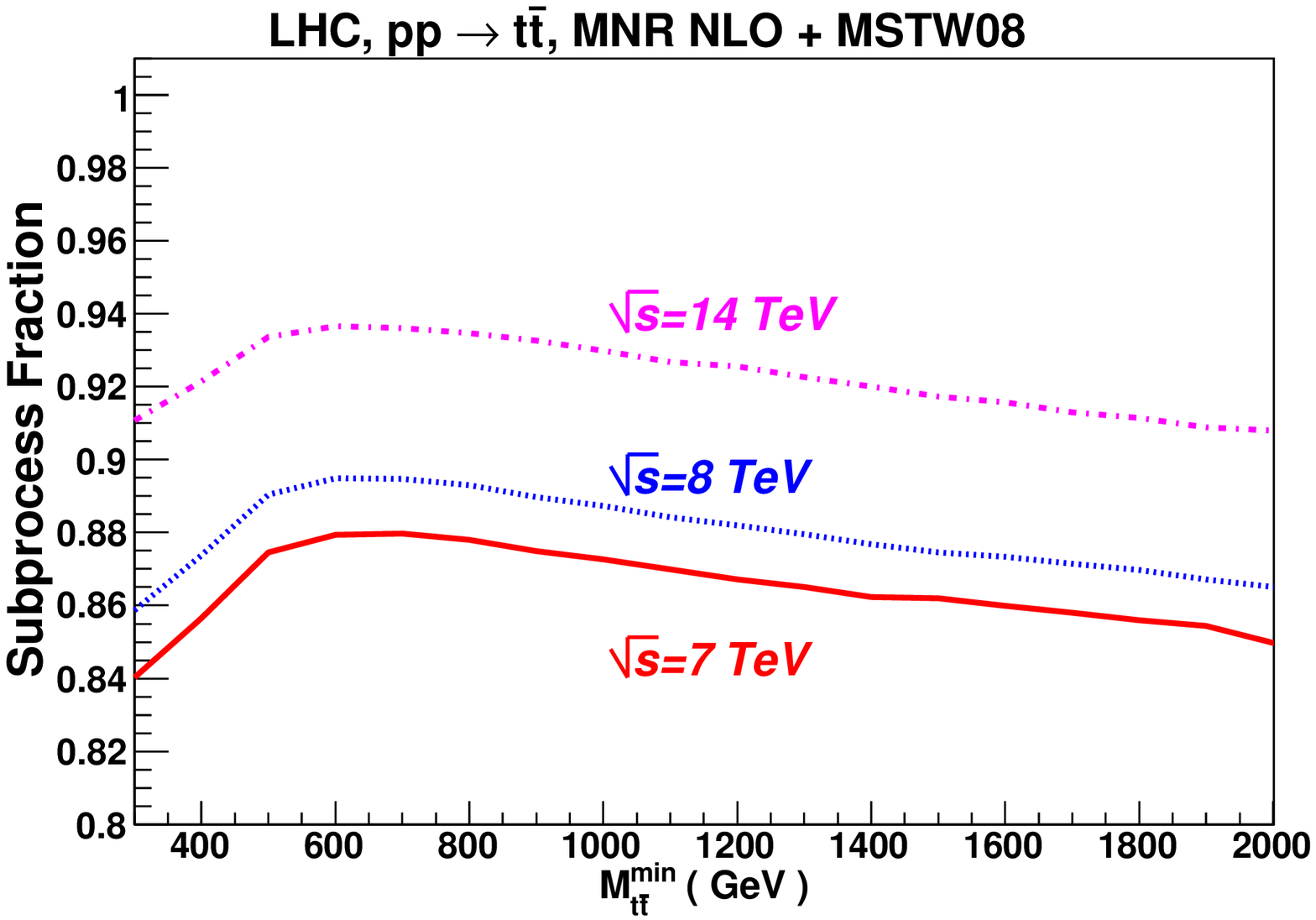 }
\epsfig{width=0.49\textwidth,figure=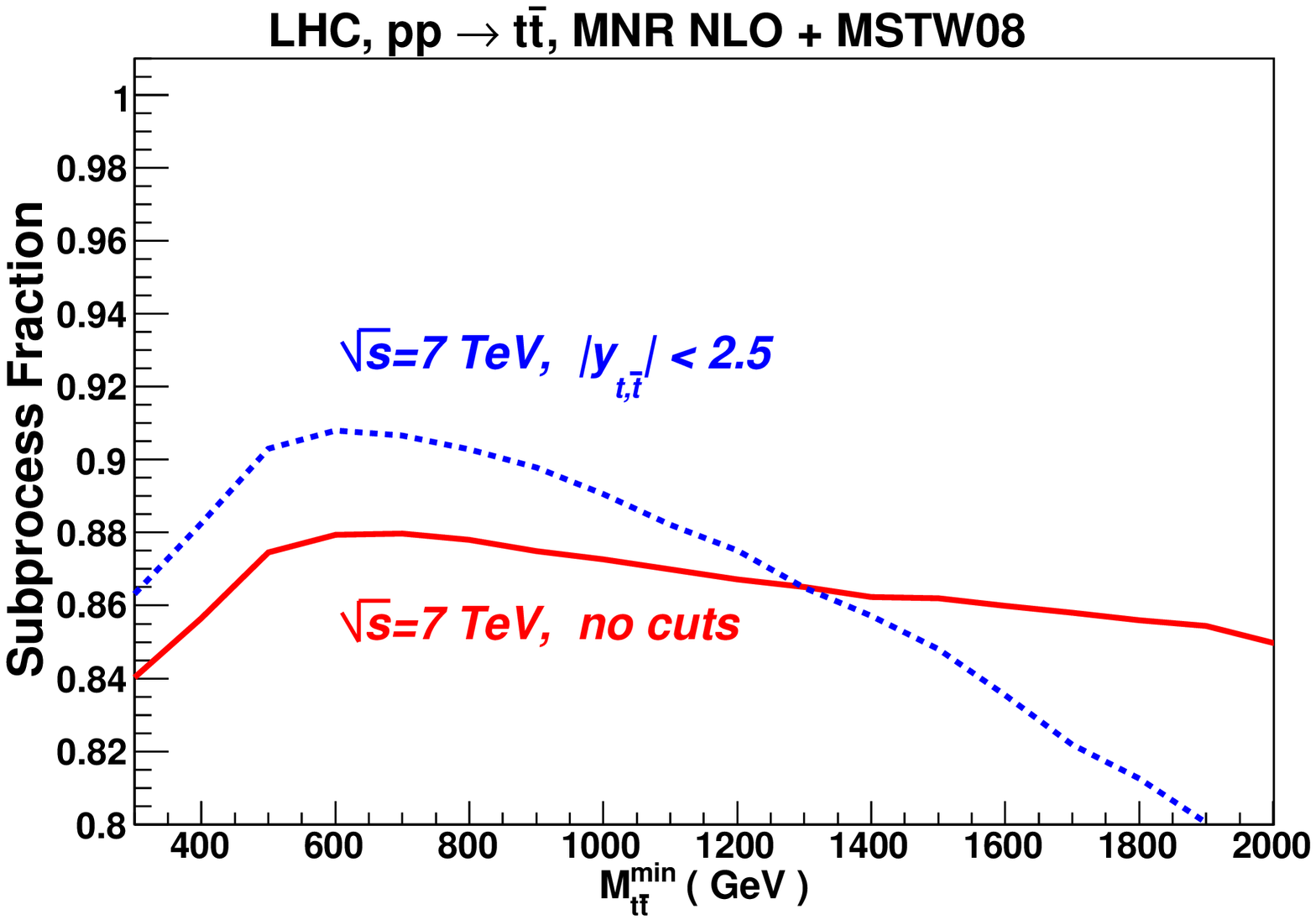 }
   \caption{\small Initial state composition for  $\ttbar$ final
     states, for events with the invariant mass of the
top quark pair above
a certain threshold, $M_{tt}>M^{\rm min}_{tt}$, determined
at NLO with the MNR code. The
   left plot shows results for fully inclusive production, at
   different beam energies. The right plot, at $\sqrt{s}=7$~TeV,
   compares the $gg$ fraction of fully inclusive final states with
   the fraction in events with $\vert y_t\vert,\vert y_{\bar t}\vert<2.5$.}
   \label{fig:topfractions}
\end{figure}

Let's consider first high mass $\ttbar$ production.
In this case, the initial state is
dominated by $gg$ fusion. This is shown in
Fig.~\ref{fig:topfractions}, where the left plot gives the fraction of
events originated by gluon-gluon collisions, as a function of the
minimum value $M_{tt}^{\rm min}$ of the
$\ttbar$ invariant mass $M_{tt}$, and for different beam energies. The
calculation has been done at NLO with the MNR code~\cite{Mangano:1991jk}
and the MSTW08 PDFs. 
We remark that this fraction is largely
constant, over a wide range of $M_{tt}$, in spite of the fact that the
$gg$ luminosity decreases with $M_{tt}$ faster than the $q\bar q$
luminosity. The reason for this behavior is that, while
$\hat\sigma_{q\bar{q}}(\ttbar) \sim 1/M^2_{tt}$, the $t$-channel quark
exchange in the $gg\to \ttbar$ sub-process leads to
$\hat\sigma_{gg}(\ttbar) \sim \log(M^2_{tt})/M^2_{tt}$. We notice that
this behavior remains qualitatively true even requiring the top
quarks to be produced in the central rapidity region $\vert
y_{t,\bar{t}} \vert < 2.5$, as shown in the right plot of
Fig.~\ref{fig:topfractions}.

%%%%%%%%%%%%%%%%%%
%%%%%%%%%%%%%%%%%

\begin{figure}[htb]
   \centering
\epsfig{width=0.45\textwidth,figure=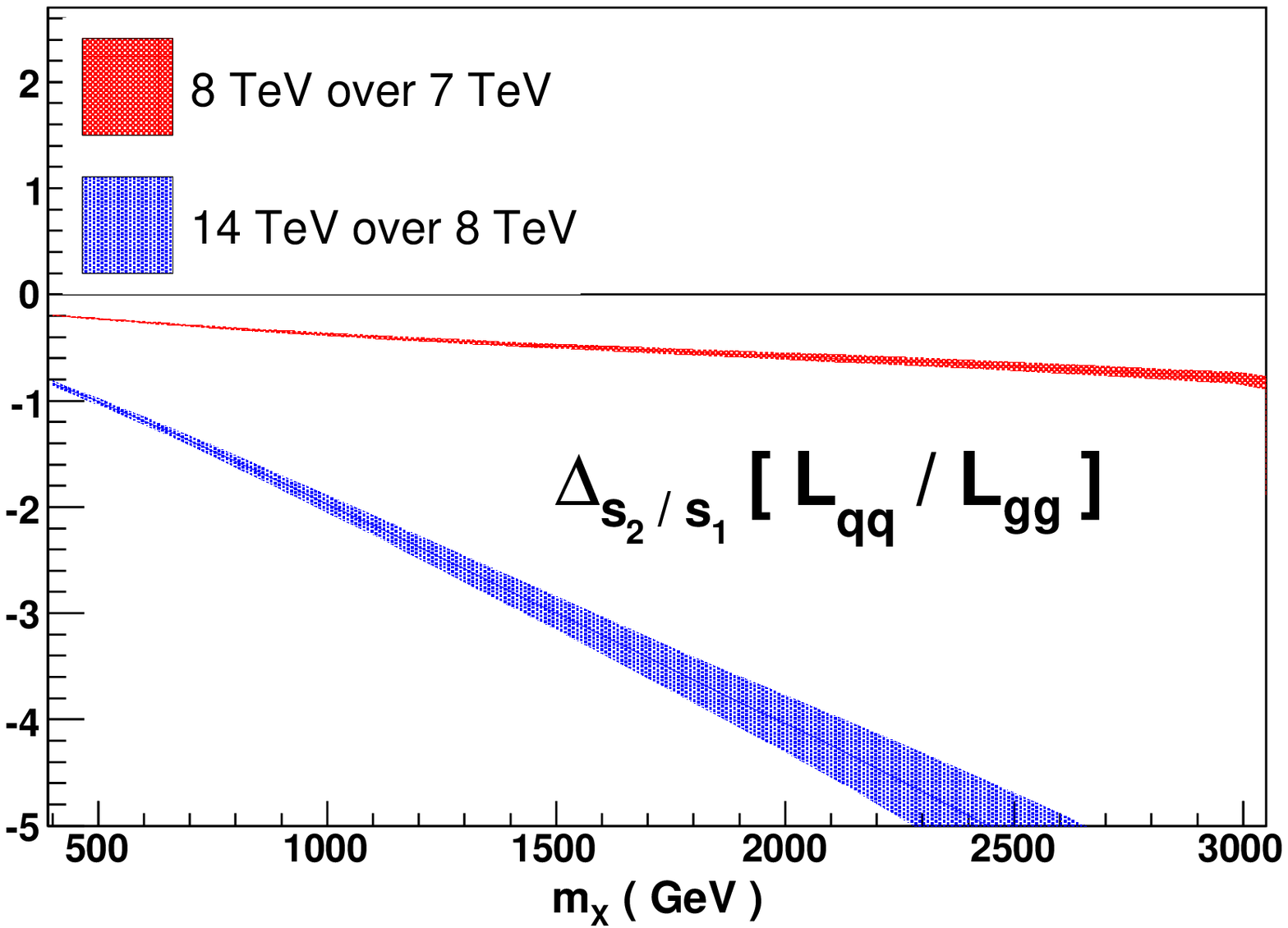 }
\epsfig{width=0.45\textwidth,figure=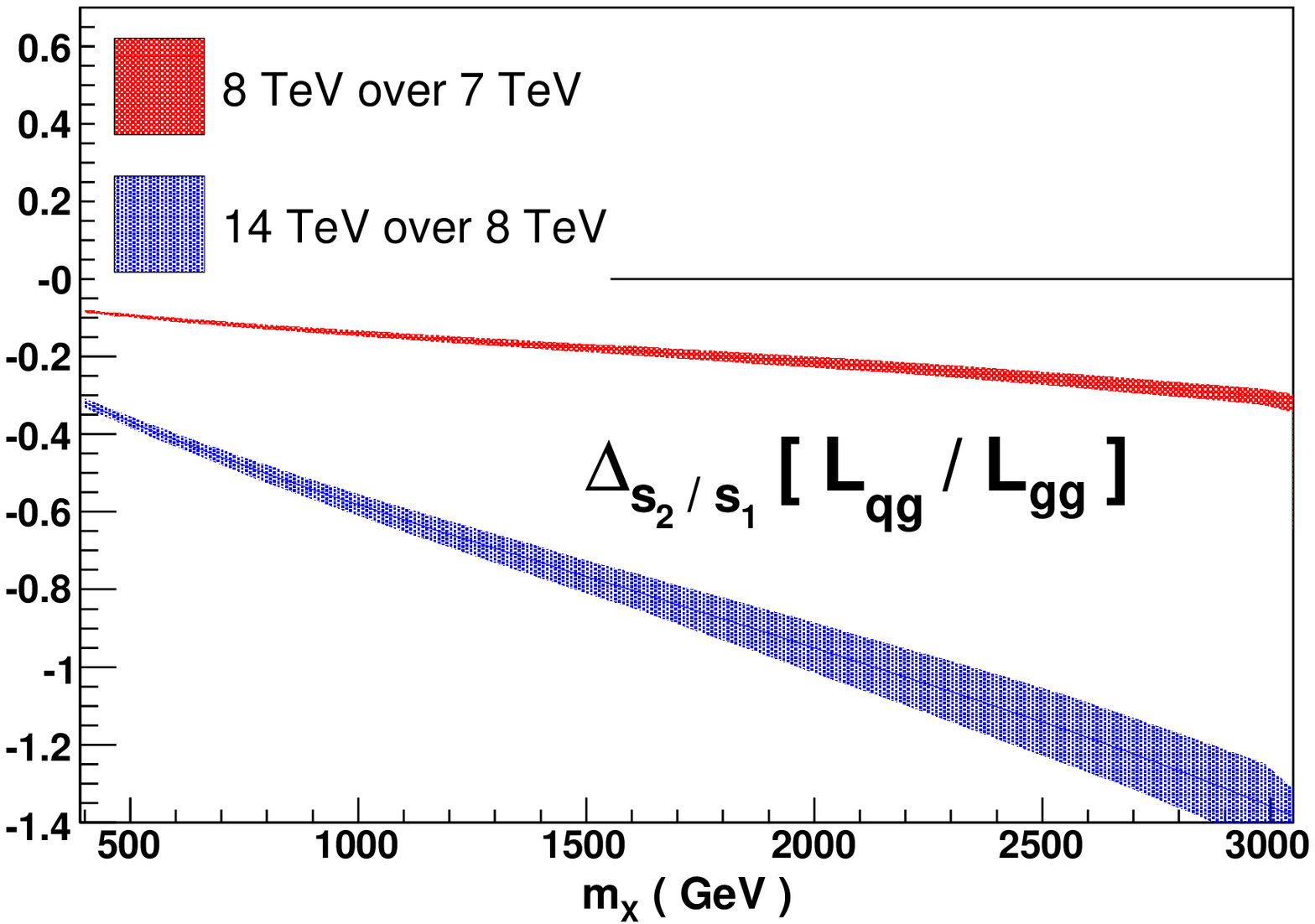 }
\epsfig{width=0.45\textwidth,figure=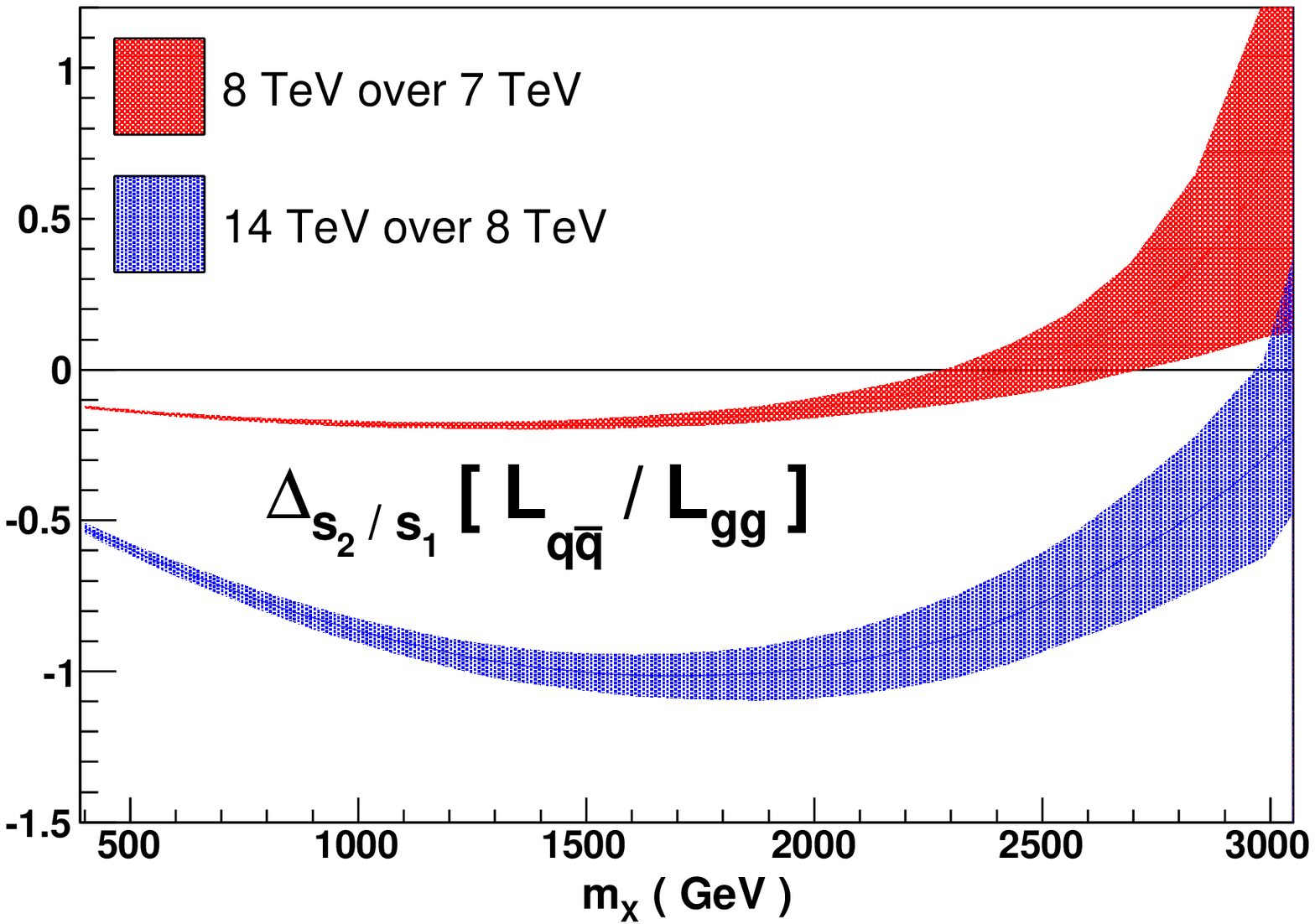 }
   \caption{\small The double ratio of luminosities between different
LHC beam energies, Eq.~(\ref{eq:deltamaster}) relevant for high
mass top quark pair production. The bands correspond to the
1--sigma PDF uncertainties.}
   \label{fig:lumirat-delta-gg}
\end{figure}

A possible BSM contribution to $\ttbar$ production driven by initial
states other than $gg$, therefore, would contribute to a deviation
from the SM energy scaling of the cross section ratio,
as dictated by Eq.~(\ref{eq:rmaster}).
For example, in the particular case of a BSM contribution to
$\sigma(\ttbar)$ due to $q\bar{q}$ initial states, 
as in the case of a $Z'$ vector boson, the deviation from the
SM scaling would be
\be
\frac{\sigma^{\rm BSM}_{\ttbar}(E_1)}{\sigma^{\rm SM}_{\ttbar}(E_1)} \; \Delta_{E_1/E_2} \left[
\frac{{\cal L}^{q\bar{q}}(M)}{{\cal L}^{gg}(M)}
\right] 
\ee
The values of the double ratio of the $q\bar{q}$ over $gg$
luminosities at
different energies, Eq.~(\ref{eq:deltamaster}), is shown 
in Fig.~\ref{fig:lumirat-delta-gg},
for ratios of 8 over 7 TeV and 14 over 8 TeV luminosities, computed
again from the NNPDF2.1 NNLO PDF set.

From Fig.~\ref{fig:lumirat-delta-gg}, is clear that
for example for a BSM contribution initiated by $q\bar{q}$,
the enhancement factor due to the different scaling
with the energy  Eq.~(\ref{eq:deltamaster})
could be $\mathcal{O}(1)$ in most of the TeV region for the 14 over 8 TeV
ratios. Given that
the systematics in the cross section ratio for top quark pair
production at large $\ttbar$ masses is 2-4\% at most (and likely
to be improved soon), the measurement of this cross section
ratio between 14 and 8 TeV should be sensitive to BSM contributions with
$\sigma^{\rm BSM}_{\ttbar}/\sigma^{\rm SM}_{\ttbar}$ well below 10\%.
For the 8 over 7 ratio
the enhancement factor is smaller, but so is the theoretical
systematics. This probe is therefore more sensitive to BSM effects
than the measurement at a fixed beam energy (unless of course one
considers trivially clear BSM signatures such as mass peaks).

For completeness we also show in Fig.~\ref{fig:lumirat-delta-gg} the
evolution of the luminosity ratios of $qq$ over $gg$, and $qg$ over
$gg$ initial states. We see that in this case 
there could be a sizable suppression of the cross section
ratios for a process whose BSM contribution is quark--quark initiated.
Indeed, for 14 over 8 TeV ratios the enhancement factor
can be up to  $\mathcal{O}(5)$ for $m_X \sim 2$ TeV.
This means that the ratio of high mass  $\ttbar$ cross sections
between 14 and 8 TeV  can be up to five times
more sensitive to BSM $qq$ initiated processes that the absolute
cross section, with the cross section prediction
and measurement being rather
more precise both theoretically and experimentally.

%%%%%%%%%%%%%%%%%%%%%%%
%%%%%%%%%%%%%%%%%%%%

\begin{figure}[t]
  \centering
\epsfig{width=0.49\textwidth,figure=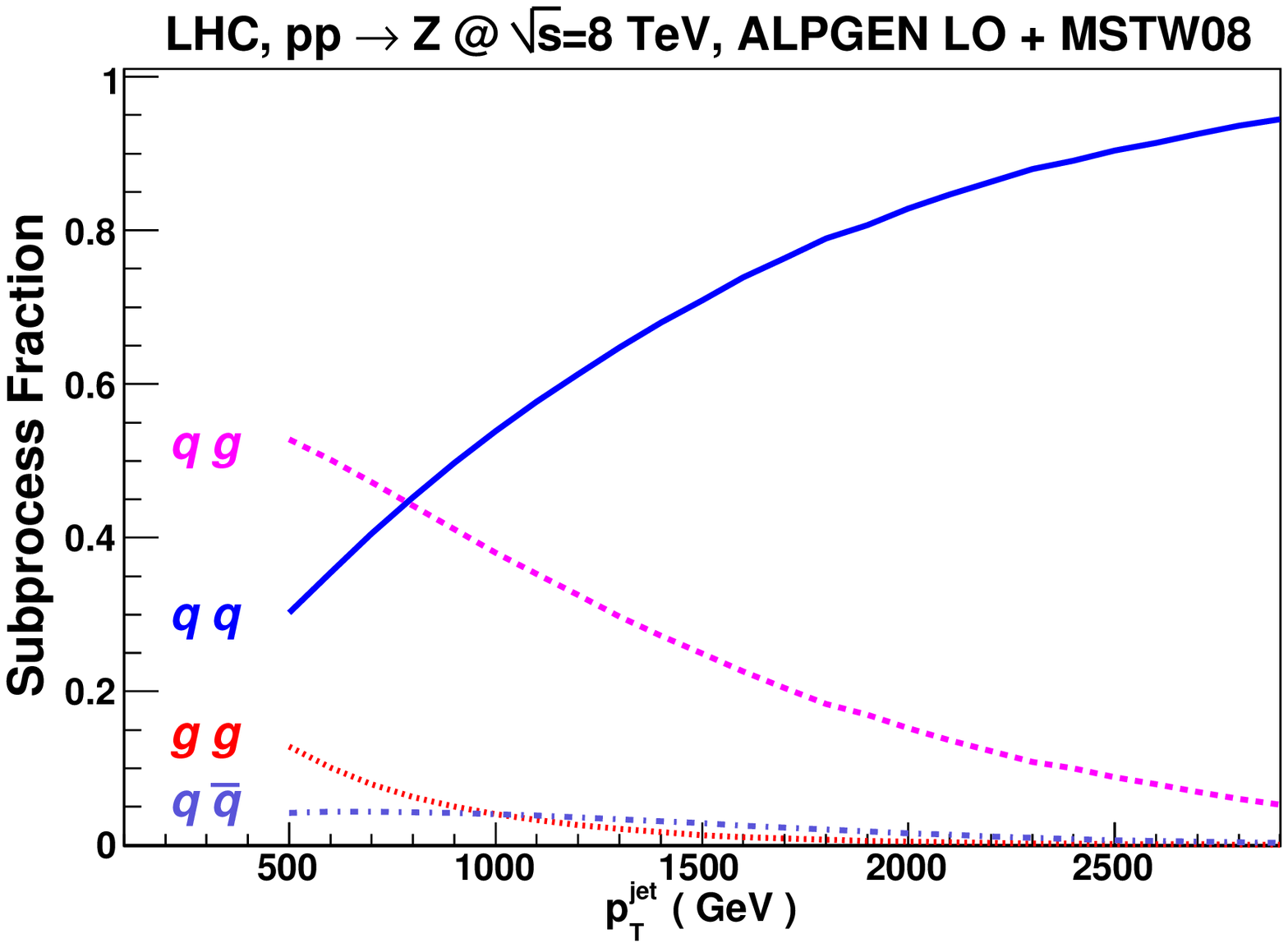}
\epsfig{width=0.49\textwidth,figure=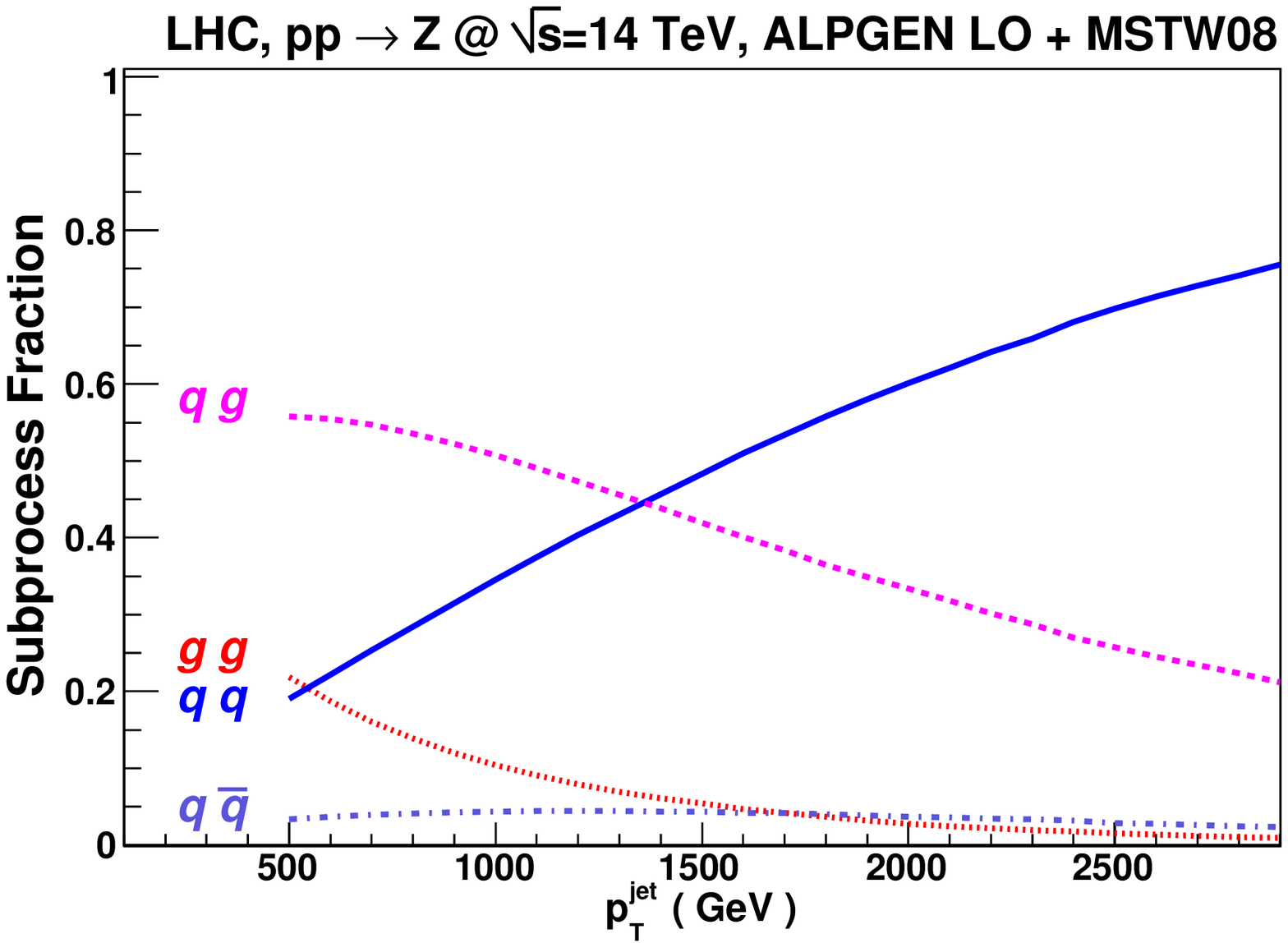}
   \caption{\small Initial state subprocess fraction for jet final
     states for 8 TeV (left plot) and 14 TeV (right plot)
jet production, as a function of the $p_T$ of the jet.
The computation has been done with ALPGEN at LO. The
decomposition of partonic subprocesses is the same 
as in Eqns.~(\ref{eq:lumi1}--\ref{eq:lumi4}).
 The decomposition is very similar between 7 TeV and
8 TeV and thus the 7 TeV case is not shown here.}
   \label{fig:jetfractions}
\end{figure}

The second illustrative example that we have considered is
inclusive jet production.
In the case of inclusive jet spectra, the dominant initial state
composition varies depending on the jet $p_T$. We calculated this, 
with {\small ALPGEN}~\cite{Mangano:2002ea}, at
leading-order, which is sufficient for our qualitative discussion.
The results are shown in
fig.~\ref{fig:jetfractions}, which gives the contributions of the
$gg$, $qg$, $q\bar{q}$ and of the quark-quark 'elastic' channel,
$qq^{(')}\to qq^{(')}$. At large
$p_T$, the latter largely dominates, but there is a large range where
$qg$ is also important. The
sensitivity of energy ratios of jet spectra to new physics can
therefore only be established on a case-by-case basis.

%%%%%%%%%%%%%%%%%%%%%%%%%%
%%%%%%%%%%%%%%%%%%%%

\begin{figure}[htb]
   \centering
\epsfig{width=0.45\textwidth,figure=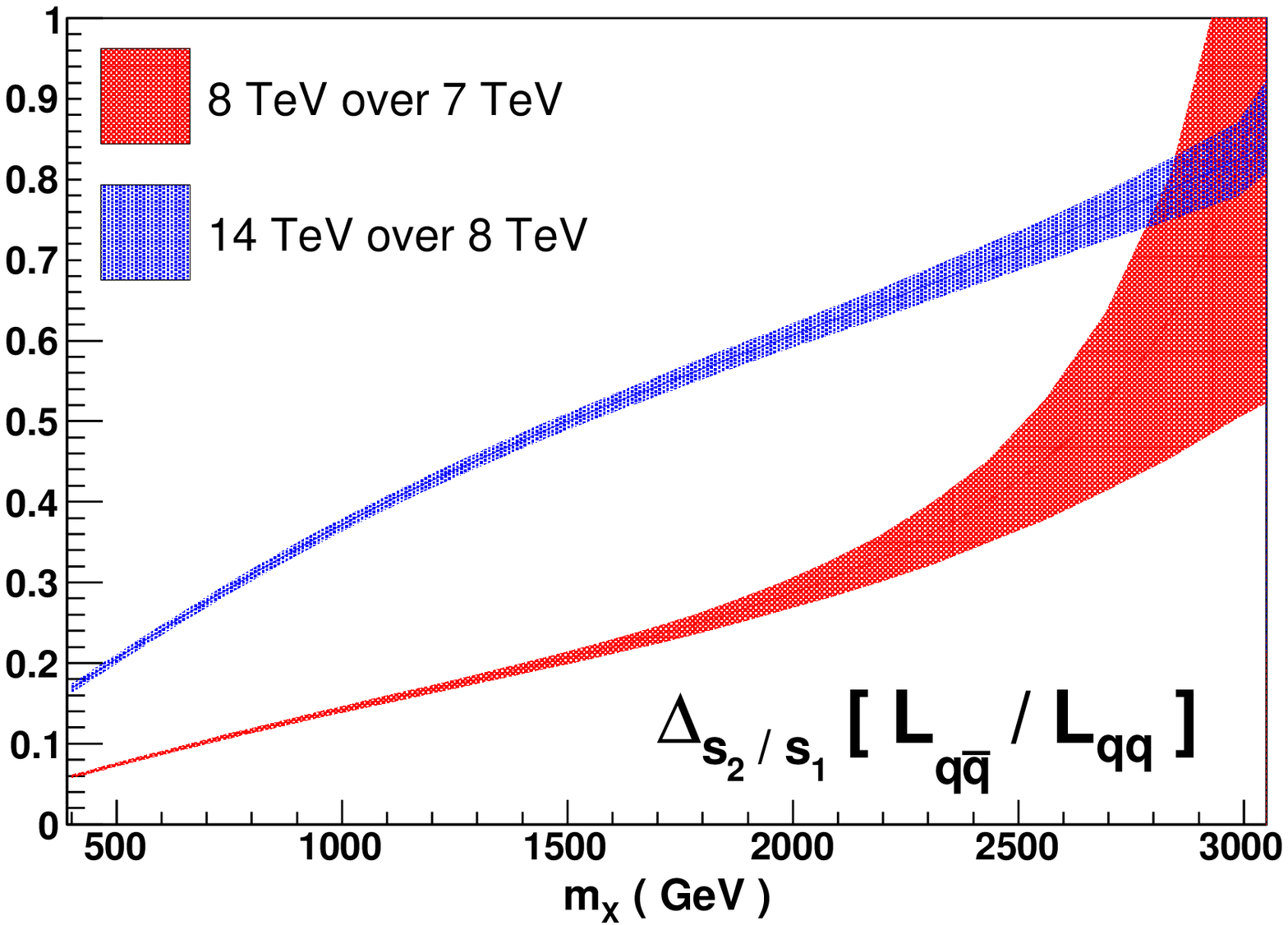 }
\epsfig{width=0.45\textwidth,figure=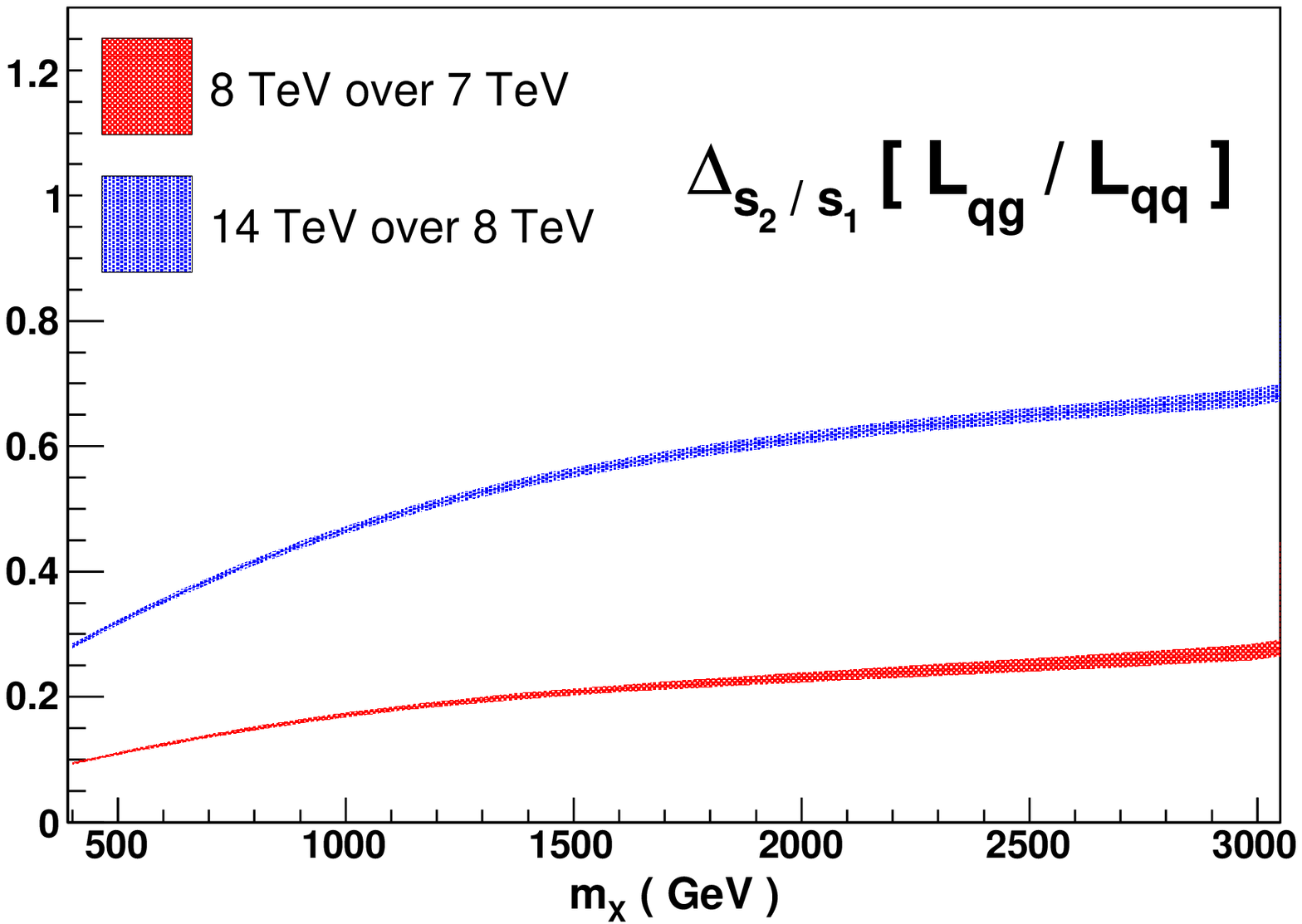 }
\epsfig{width=0.45\textwidth,figure=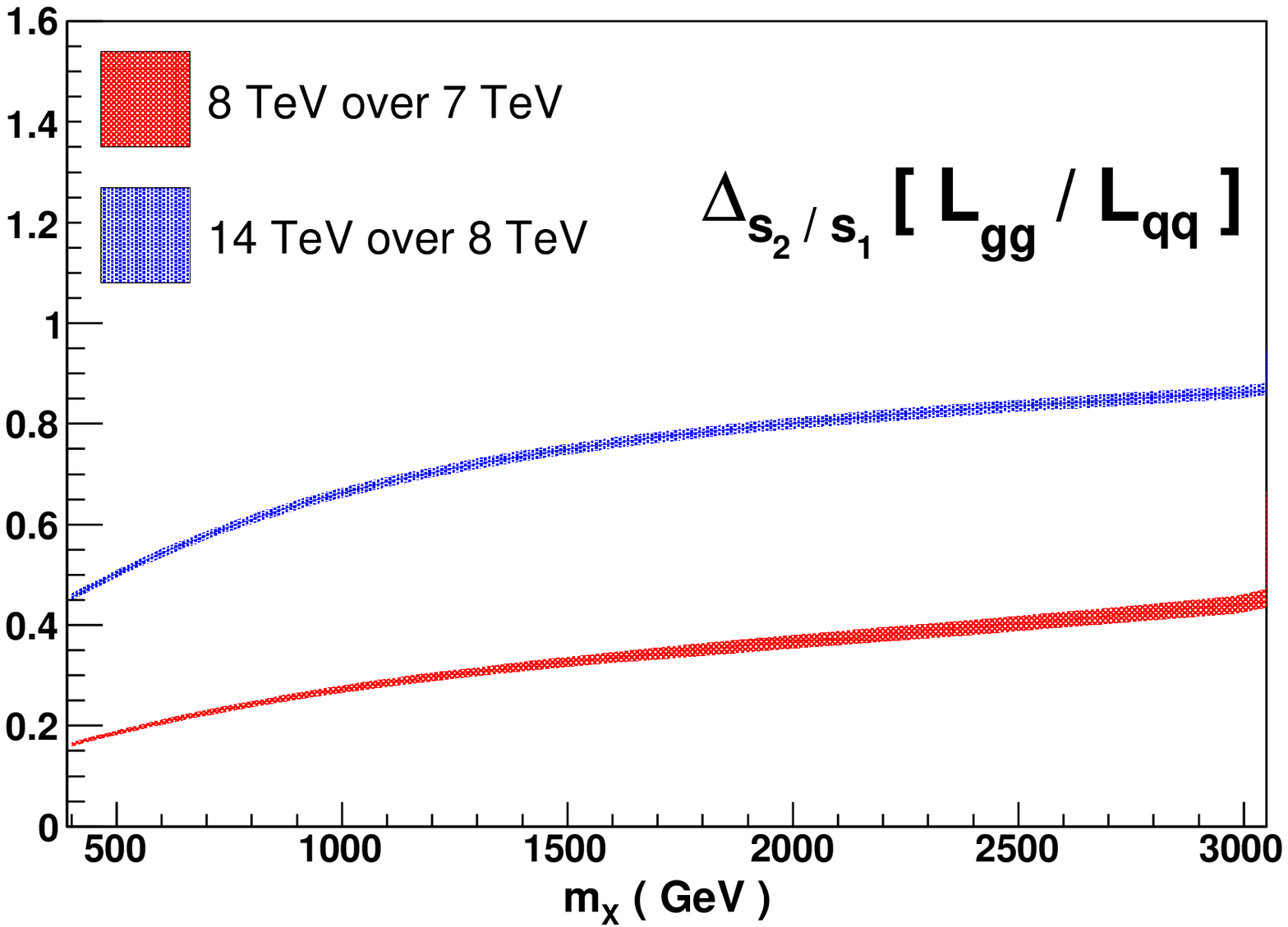 }
   \caption{\small The double ratio of luminosities between different
LHC beam energies, Eq.~(\ref{eq:deltamaster}) 
relevant for BSM searches in high $p_T$ inclusive
jet production. The bands correspond to the
1--sigma PDF uncertainties. }
   \label{fig:lumirat-delta-qq}
\end{figure}

A possible BSM contribution to inclusive jet production
at high--$p_T$ driven by a
$q\bar{q}$, $gg$ or $qg$ initial state, therefore, 
would contribute to a deviation
from the SM energy scaling as dictated from Eq.~(\ref{eq:rmaster}).
The values of the double luminosity ratio
Eq.~(\ref{eq:deltamaster}) for $q\bar{q}$, $gg$ and
$qg$ luminosities over
$qq$ luminosity are represented in Fig.~\ref{fig:lumirat-delta-qq},
for ratios of 8 over 7 TeV and 14 over 8 TeV luminosities. It is clear
that in this case the enhancement of possible BSM contributions
is more moderate but still appreciable, reaching
 $\mathcal{O}(1)$ at large masses for a $q\bar{q}$ or $gg$--initiated
BSM contribution. Thus the measurement of high--$p_T$ jet cross
section ratios at different LHC energies, if precise enough, could
provide a competitive search strategy for BSM scenarios that lead
to the same jet final state.

%%%%%%%%%%%%%%%%%%%%%%%
%%%%%%%%%%%%%%%%%%%%

\begin{figure}[t]
  \centering
\epsfig{width=0.49\textwidth,figure=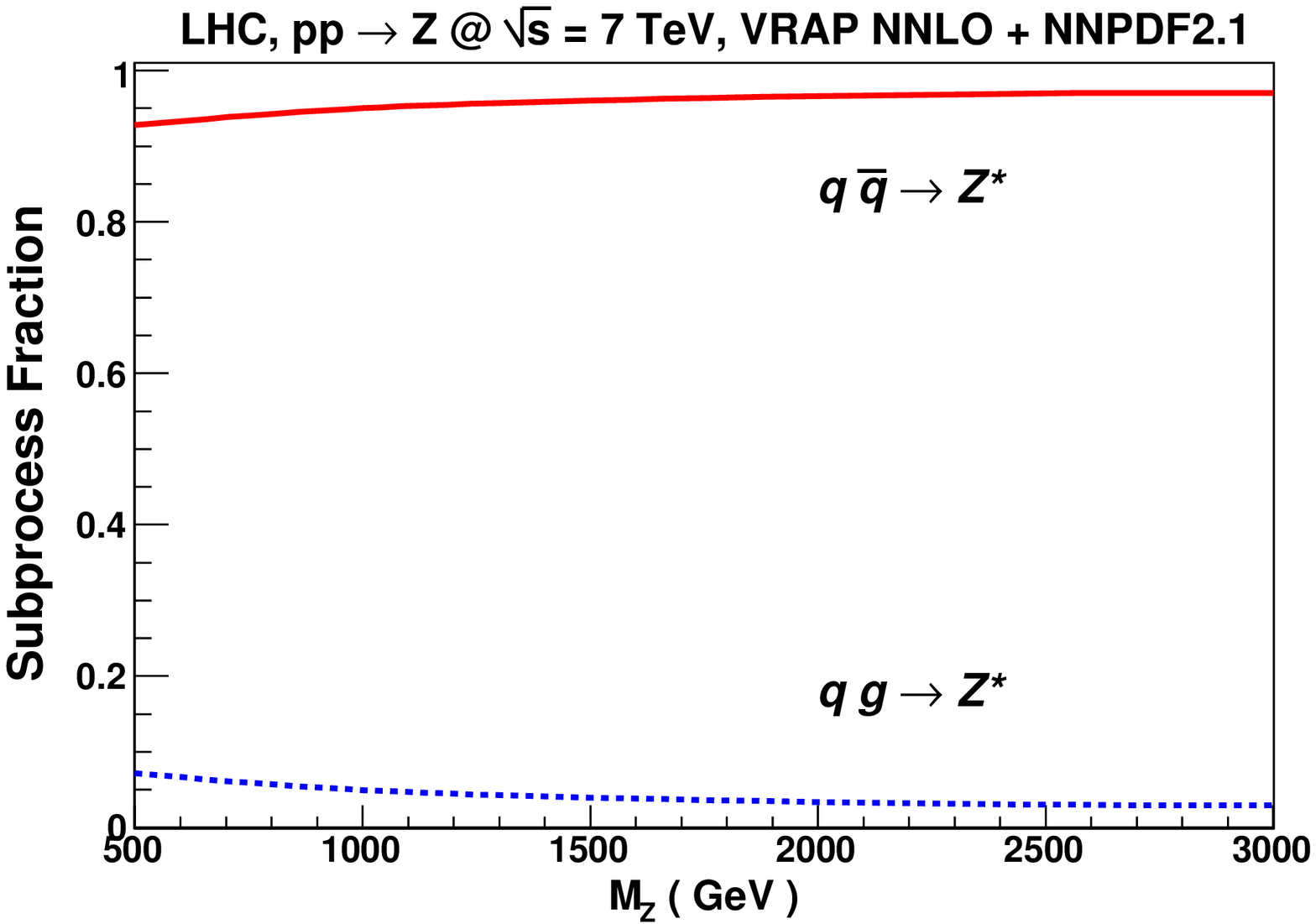}
\epsfig{width=0.49\textwidth,figure=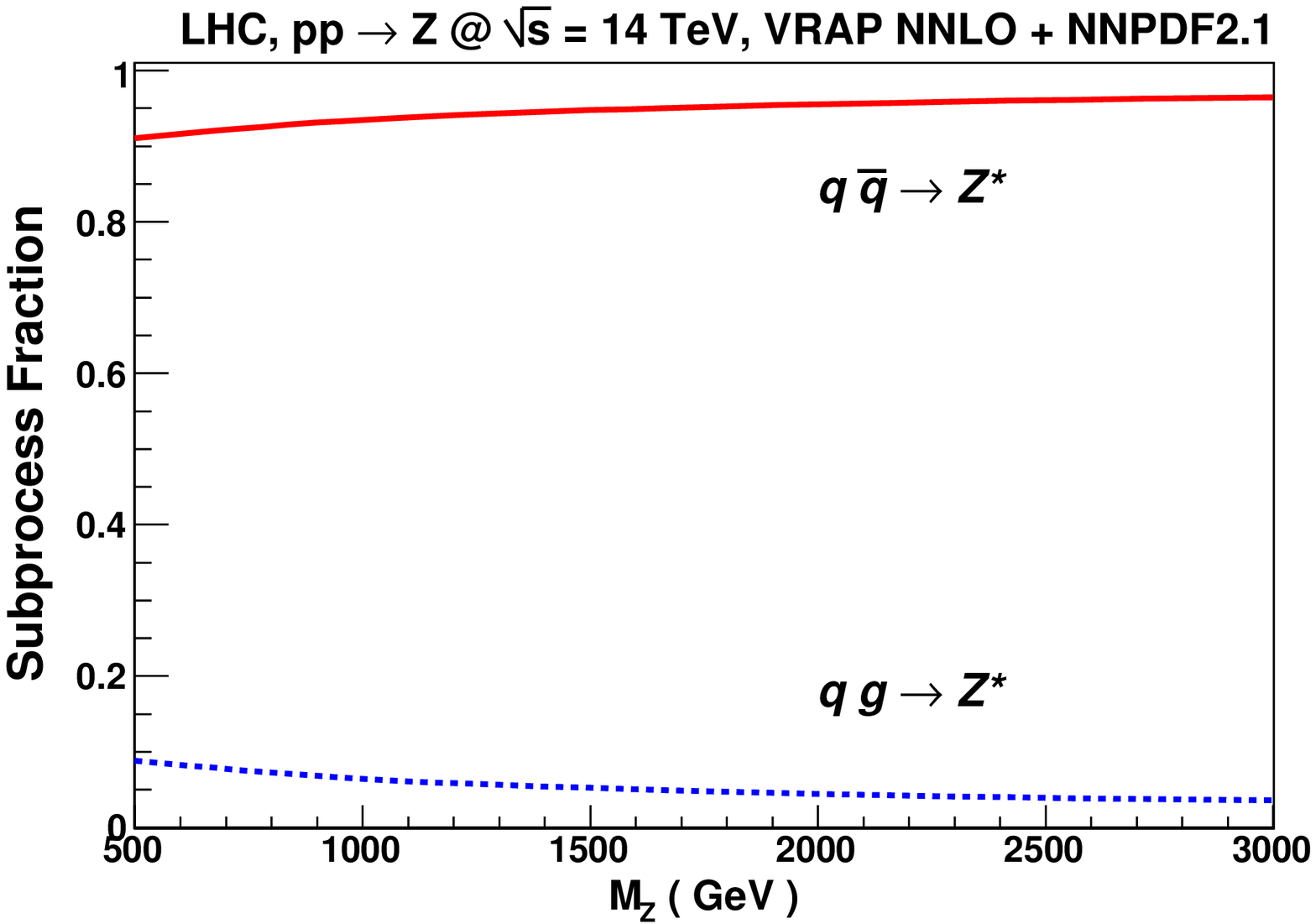}
   \caption{\small Initial state composition for
high mass Standard Model $Z$ boson production for
7 TeV (left plot) and 14 TeV (right plot) LHC,
as a function of the invariant mass of the off-shell
$Z$ boson.
The computation has been done with VRAP at NNLO,
with NNPDF2.1 as input. 
 The decomposition is very similar between 7 TeV and
8 TeV and thus the latter case is not shown here.}
   \label{fig:zfractions}
\end{figure}

The final example is high mass off-shell $Z$ boson
production.  The initial state composition for 
high mass Standard Model $Z$ boson production at the
LHC for
7 TeV and 14 TeV,
as a function of the invariant mass of the off-shell
$Z$ boson, is shown in Fig.~\ref{fig:zfractions}.
The computation has been done with {\tt Vrap} code at NNLO,
with NNPDF2.1 as input.
It is clear that the quark-antiquark scattering dominates
at all masses, and the $qg$ contamination is reduced to a few
percent.

%%%%%%%%%%%%%%%%%%%%%%%%%%
%%%%%%%%%%%%%%%%%%%%

\begin{figure}[htb]
   \centering
\epsfig{width=0.45\textwidth,figure=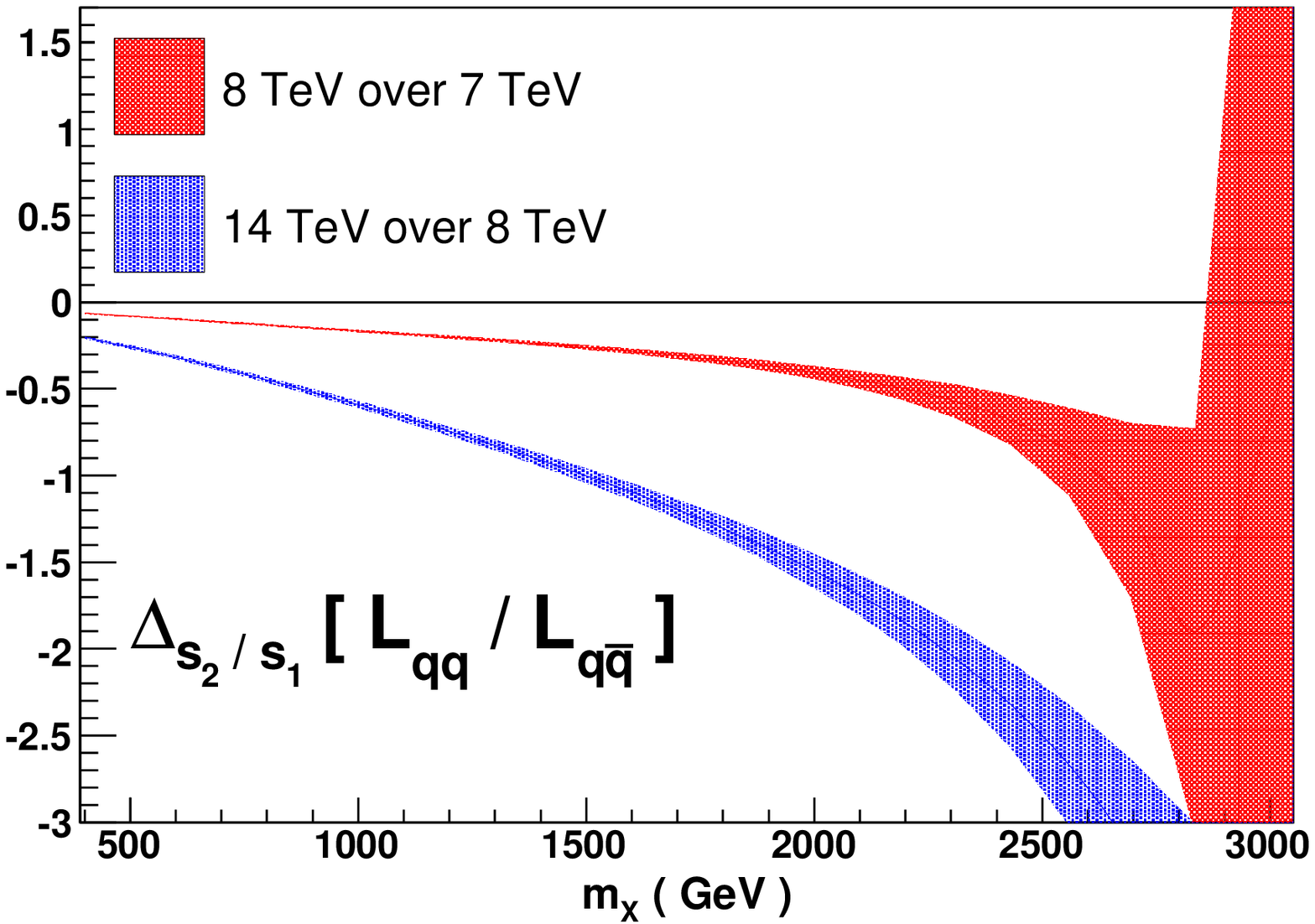 }
\epsfig{width=0.45\textwidth,figure=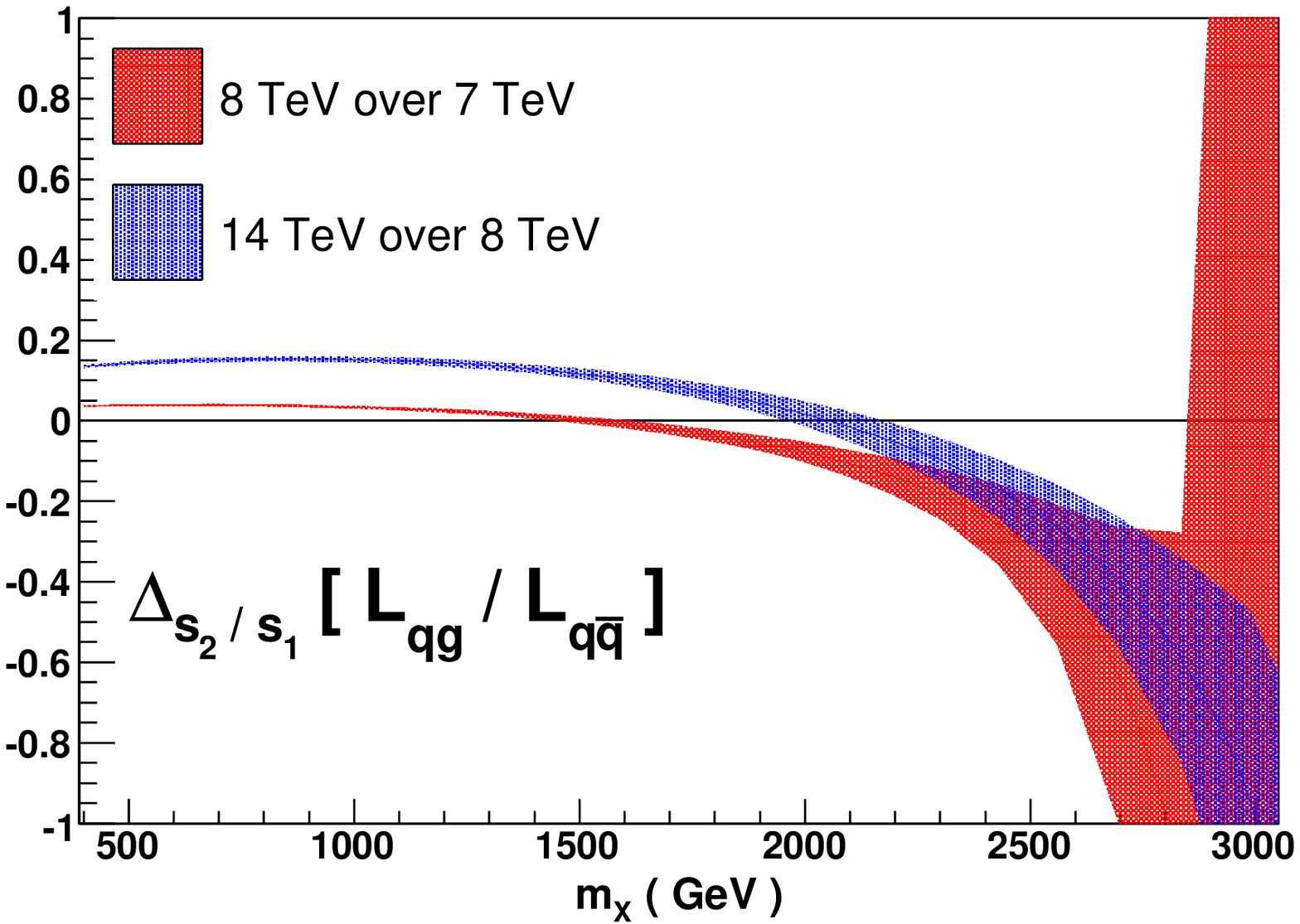 }
\epsfig{width=0.45\textwidth,figure=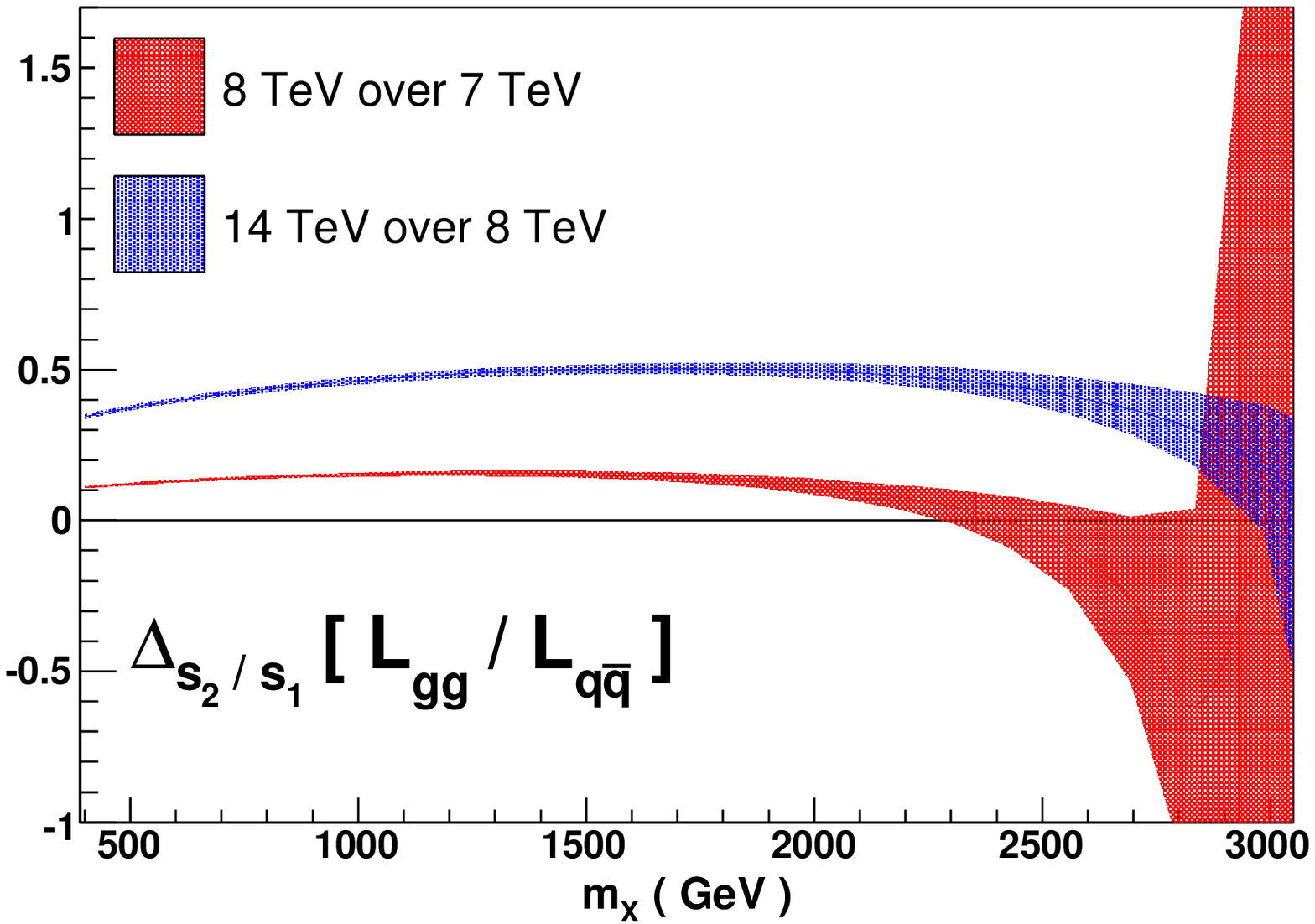 }
   \caption{\small The double ratio of luminosities between different
LHC beam energies, Eq.~(\ref{eq:deltamaster}) for relevant
for BSM searches in high-mass off shell $Z$ boson
production. The bands correspond to the
1--sigma PDF uncertainties.}
   \label{fig:lumirat-delta-qqb}
\end{figure}

The double ratios  between different
LHC beam energies
 relevant to this case are shown in 
Fig.~\ref{fig:lumirat-delta-qqb}.  For 8 over 7 TeV
ratios the scaling with center of mass energy is similar
for all luminosities and thus the enhancement of BSM
signals is small, except possibly for the
highest final state masses where PDF uncertainties
explode. For 14 over 8 TeV cross section ratios, instead, the larger
lever arm leads to a more important enhancement factor
Eq.~(\ref{eq:deltamaster}). For example, this
can be $\mathcal{O}(2)$
for BSM $qq$--initiated contributions, and
$\mathcal{O}(0.5)$
for BSM $gg$--initiated contributions.

%%%%%%%%%%%%%%%%%%%%%%%%%%%%%%%
%%%%%%%%%%%%%%%%%%%%

\section{Conclusions}

We highlighted in this paper the potential interest in precise
measurements of ratios and double ratios of cross-sections 
at different LHC energies. The theoretical precision with which such
quantities can be predicted is very high. It can be better than
$10^{-3}$ for electroweak processes, but it is below the percent level even in
the case of inclusive $\ttbar$ production, an no larger than
a few percent for TeV observables like high mass  $\ttbar$
and jet production. Residual theoretical
systematics are typically dominated by PDF uncertainties. When these
are large enough that the experimental measurements are sensitive to
them, the information can be used to improve the knowledge of large--$x$
PDF, a region which is crucial for high mass BSM searches. 
When these are too small, the relevant ratio can be used as a
precise calibration of the relative luminosity of runs at different
energies or between different experiments. 
We also showed that these measurements could expose the
presence of small BSM contributions, which may be smaller than the
theoretical and experimental systematics at a single energy, but which
can alter the energy evolution of the relevant cross sections by a
amount larger than the estimated uncertainty and thus be within the
reach of the LHC experiments.

The experimental measurements of these ratios and double ratios with
the required precision is certainly very challenging, and will require
dedicated analyses. Trivial issues, such as generating large enough
Monte Carlo statistics to carry out the necessary studies, may also turn
out to be possible obstacles. We hope nevertheless that the potential interest
in these results, as documented in this note, 
is compelling enough to stimulate more realistic
assessments by the experimental collaborations.
\vspace{0.5cm}

{\bf\noindent  Acknowledgments \\}
The research of J.~R.
is supported by a Marie Curie Intra--European Fellowship
of the European Community's 7th Framework Programme under contract
number PIEF-GA-2010-272515. This work was performed in the framework
of the ERC grant 291377, ``LHCtheory: Theoretical predictions and
analyses of LHC physics: advancing the precision frontier''. 
We acknowledge the help of K. Rabbertz with
{\tt FastNLO}, J. Gao with {\tt MEKS}  and A. Mitov with {\tt top++} .

\bigskip

%\bibliography{DoubleRatiosTeX}
% \input{DoubleRatiosTeX.bbl}

%%%%%%%%%%%%%%%%%%%%%%%%%%%%%%%%%%%%%%%%%%%%%%%

\end{document}